\documentclass[twocolumn,amssymb, nobibnotes, aps, prd, superscriptaddress]{revtex4-1}
\usepackage{graphicx}
\usepackage{dcolumn}
\usepackage{bm}
\usepackage{comment}
\usepackage{amsthm}
\usepackage{tikz-cd}
\usepackage{mathdots}
\usepackage{nccmath}
\usepackage{nicematrix}
\usepackage{multirow}    
\usepackage{array}       
\usepackage{tikz}
\usetikzlibrary{decorations.markings}
\usetikzlibrary{shapes.geometric}
\pgfdeclarelayer{edgelayer}
\pgfdeclarelayer{nodelayer}
\pgfsetlayers{edgelayer,nodelayer,main}
\usepackage{algorithm}
\usepackage{algpseudocode}
\usepackage{float}
\theoremstyle{definition}
\newtheorem{theorem}{Theorem}[section]
\newtheorem{corollary}{Corollary}[theorem]

\newtheorem{claim}{Claim}[section]
\newtheorem{definition}{Definition}[section]

\newtheorem{example}{Example}[section]
\newcommand\quotient[2]{
	\mathchoice
	{
		\text{\raise1ex\hbox{$#1$}\Big/\lower1ex\hbox{$#2$}}%
	}
	{
		#1\,/\,#2
	}
	{
		#1\,/\,#2
	}
	{
		#1\,/\,#2
	}
}
\newcommand*{\addheight}[2][.5ex]{%
	\raisebox{0pt}[\dimexpr\height+(#1)\relax]{#2}%
}
\newcommand\rank{rank\:}

\usetikzlibrary{patterns}
\tikzstyle{blackdot}=[fill=black, draw=black, shape=circle]
\tikzstyle{bluedot}=[fill=white, draw=blue,line width=1pt, shape=circle]
\tikzstyle{none}=[]
\tikzstyle{arrow}=[stealth-stealth,draw=black, line width=1pt]
\tikzstyle{arrow2}=[-stealth,draw=black, line width=1pt]
\tikzstyle{greenedge}=[-, draw=black!40!green,dashed,line width=1pt]
\tikzstyle{blueedge}=[-, draw=blue,line width=1pt]
\tikzstyle{blackedge}=[-, draw=black,line width=0.75pt]
\tikzstyle{bluearrow2}=[-stealth,draw=blue, line width=1pt]

\begin{document}


\title{Design Principles for Realizable Discrete Surface Embeddings in Physical Systems}
\thanks{This paper is supported by NSF EFRI Project 1935294}%

\author{Kyungeun Kim}
\email{kkim10@syr.edu}
 \affiliation{Physics Department, Syracuse University}
\author{Christian D. Santangelo}%
 \email{cdsantan@syr.edu
}
\affiliation{%
Physics Department, Syracuse University}%

\date{\today}

\begin{abstract}
The isometric embedding of surfaces in three-dimensional space is fundamental to various physical systems, from elastic sheets to programmable materials. While continuous surfaces typically admit unique solutions under suitable boundary conditions, their discrete counterparts-represented as networks of vertices connected by edges-can exhibit multiple distinct embeddings for identical edge lengths. We present a systematic approach to constructing discrete meshes that yield a controlled number of embeddings. By analyzing the relationship between mesh connectivity and embedding multiplicity through rigidity theory, we develop criteria for designing meshes that minimize solution multiplicity. We demonstrate computational methods based on local matrix operations and trilateration techniques, enabling practical implementation for meshes with approximately a thousand vertices. Our analysis provides both theoretical bounds on the number of possible embeddings based on Bézout's theorem and practical guidelines for mesh construction in physical applications. Through numerical simulations, we show that this approach achieves comparable accuracy to traditional minimization methods while offering computational advantages through sequential computation. Importantly, we demonstrate that in cases where a unique smooth solution exists, local fluctuations in reconstructed shapes derived from the computational grid can serve as indicators of insufficient geometric constraints. This work bridges the gap between discrete and continuous embedding problems, providing insights for applications in 4D printing, mechanical meta-materials, and deployable structures.
\end{abstract}
\maketitle

\section{Introduction}

Networks of interconnected nodes with prescribed edge lengths appear across diverse fields, from physical and biological systems to machine learning architectures \cite{Ravasz2003, Newman2003}. These networks often represent fundamental relationships: interatomic distances in molecular structures \cite{Liberti2014}, resistive or elastic connections in training networks \cite{Poole2016}, geometric constraints in deployable structures \cite{Chen2015}, or elastic or deformable materials with networks \cite{Smith2002,Panetta2015}. A key question is this: when the lengths of the edges of a network have been prescribed, is there a natural way to embed the nodes to achieve these lengths? While this embedding problem seems straightforward, it reveals surprising complexity in practice \cite{Connelly1996, Dgeom,Jackson2009} because multiple distinct spatial configurations can satisfy identical edge-length constraints \cite{Hendrickson1992}.

This paper is specifically concerned with embedding triangular meshes that approximate surfaces into $\mathbb{R}^3$. In the continuum limit, specifying lengths on a triangular mesh is naively equivalent to the problem of the isometric embedding of a metric in $\mathbb{R}^3$ \cite{Bryant1983,Han2010, Janet1927, han2006isometric}. 
For example, in the emerging field of 4D printing, a flat, elastic film is programmed to swell or shrink according to a spatially inhomogeneous pattern, ultimately buckling into a target three-dimensional shape \cite{Klein2007}. In the limit of zero thickness, the resulting shape approximates an isometric embedding of the (mid-surface) metric \cite{Efrati2009,Lewicka2010}. While there are challenging examples -- for instance, a flat torus can be embedded in three dimensions continuously but not differentiably \cite{FlatToriEmb, gromov1973convex} -- for sufficiently well-behaved metrics with the topology of a disk, smooth embeddings of a finite area are generally feasible \cite{Janet1927}.
In practice that area can be made arbitrarily large depending on the physical constraint of avoiding self-intersections.

Since purely analytical solutions of continuous geometry are often computationally intractable, surface discretization using triangular or square meshes is a common approach. Yet, the transition from continuous to discrete representations introduces fundamental challenges in both theory and computation \cite{Bouaziz2014}. For example, Marder \textit{et al.} \cite{Marder2006} found that, counter to intuition, dramatically different configurations could satisfy identical edge constraints to numerical precision. Even though these configurations represent mathematically valid solutions for the discrete mesh, they often differ significantly from the desired continuous target state \cite{Chen2010}. Numerical papers suggest that using proper mesh combined with obtaining stable mesh solutions that approximate the continuous case requires careful initialization and iterative refinement \cite{Hotz2004,Brandt2011,Skouras2014}.

From these observations, several important questions emerge: What factors determine the number of possible configurations for a given triangulated mesh? Can we design network topologies that guarantee a unique, or at least a bounded, number of solutions? And finally, can we efficiently compute these configurations without requiring a global optimization? This paper addresses these questions through a systematic analysis of graph embedding multiplicity and introduces practical methods for constructing networks with controlled solution spaces.

Our key contributions include:
\begin{enumerate}
\item A theoretical framework for analyzing and bounding the number of possible embeddings for a given network topology
\item Design principles for constructing networks that minimize solution multiplicity
\item Demonstration of computational methods based on specific grid structures
\item A demonstration that local disorder in discretized reconstructions can serve as indicators of insufficient geometric constraints when a unique smooth solution exists
\item Practical guidelines for implementing these approaches in physical systems
\end{enumerate}

The paper is organized as follows. Section II establishes the theoretical foundations of graph embedding and its relationship to rigidity theory. Section III presents our methods for constructing networks with controlled embedding properties. Section IV details the application of computational methods to specific grid structures. Section V demonstrates the effectiveness of our approach through numerical examples, including analysis of local fluctuations as indicators of constraint insufficiency. We conclude with a discussion of implications and future directions.

\section{Shape Reconstruction from Distance Constraints}
\begin{figure}[]
    \centering
    \includegraphics[width=0.45 \textwidth]{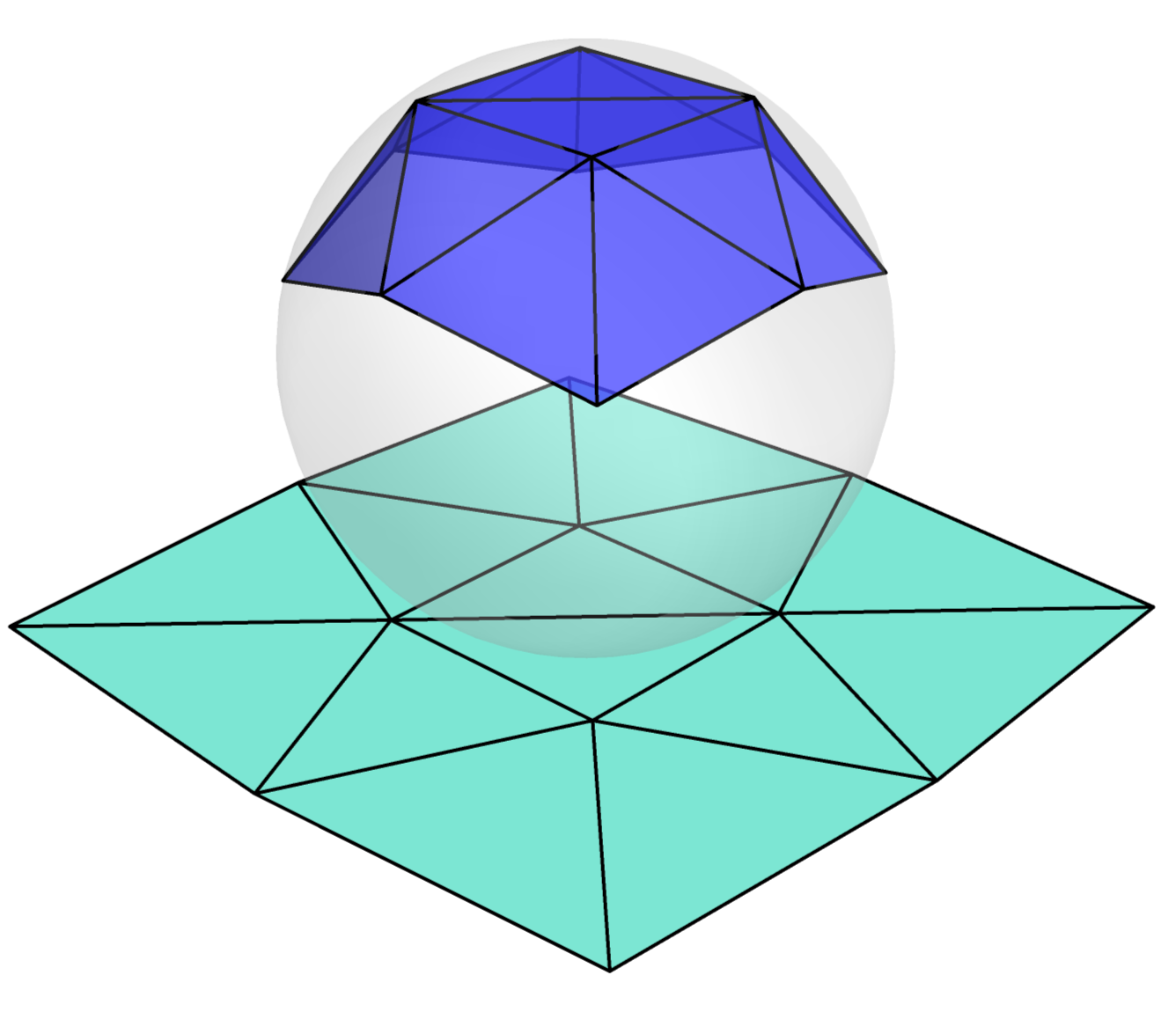}
    \caption{An illustration of a continuous surface approximated by triangulation and projected onto a plane. The triangular mesh provides a piecewise linear approximation of the smooth surface.}
    \label{fig:triangulation}
\end{figure}

In this section, we examine the differences between continuum and discrete approaches to solving isometric embedding problems. Due to computational complexity, smooth surfaces are often approximated by discretized structures such as square grids or triangulated meshes, as shown in Fig. \ref{fig:triangulation}. This discretization facilitates numerical computation of differential equations and integrals by converting analytical expressions to discrete operations. However, the geometric differences between smooth ($C^\infty$) and discretized (typically $C^1$ or $C^0$) representations lead to significant differences in embedding results when only first-fundamental form constraints (related to elasticity) are provided.

In the following subsections, we systematically develop a framework for understanding the relationship between continuous and discrete surface representations. We first explore shape reconstruction in the continuum, establishing the mathematical foundations of isometric embedding. Then, we examine how these principles translate to discrete structures, highlighting the fundamental differences that arise when moving from continuous mathematics to computational implementations. Through this analysis, we establish precise conditions under which discrete approximations can effectively represent continuous surfaces.

\subsection{Shape reconstruction in the continuum}
The question of how a physical shape emerges from prescribed distance constraints appears across diverse systems, from elastic sheets to molecular structures. This reconstruction problem has traditionally been studied through the lens of isometric embedding theory. However, to avoid technical ambiguities arising from different mathematical definitions of isometric embedding, we frame our discussion in terms of the more physically intuitive concept of shape reconstruction. Our analysis proceeds from physical motivation to mathematical framework, highlighting the parallel between continuous and discrete systems.

The behavior of a thin elastic sheet provides an instructive physical model for understanding shape reconstruction. Consider a sheet of thickness $t$ that can deform in three-dimensional space. The shape of this sheet is described by a function $\mathbf{X}(r,s)$ mapping coordinates $(r,s)$ on the sheet to positions in space. The energetic cost of deformation is given by.
\begin{equation}
\label{eq:efrati}
    E = t \int dA ~ \left( \gamma_{i j} S^{i j k l} \gamma_{k l} + t^2 m_{i j} B^{i j k l} m_{k l} \right),
\end{equation}
where each term has clear physical meaning. The first term penalizes in-plane stretching, with $\gamma_{i j}$ measuring strain within the sheet. The second term penalizes bending, with $m_{i j}$ measuring curvature. The moduli $S^{i j k l}$ and $B^{i j k l}$ are material-dependent elastic constants. Finally, the factor $t^2$ ensures that bending is cheaper than stretching as the sheet thickness goes to zero.

The strain $\gamma_{i j} = ( \partial_i \mathbf{X} \cdot \partial_j \mathbf{X} - \bar{g}_{i j} )/2$ measures how local distances in the deformed configuration (given by $\partial_i \mathbf{X} \cdot \partial_j \mathbf{X}$) deviate from the prescribed distances (encoded in the metric $\bar{g}_{i j}$). As the thickness $t$ approaches zero, energy minimization requires $\gamma_{i j} = 0$, meaning the sheet must deform without stretching. Moreover, intuitively, stretching corresponds to the first fundamental form, which encodes intrinsic distances and geodesics, while bending relates to the second fundamental form, which describes the curvature of the embedding.

This physical system motivates a central question: given a set of prescribed distances, what shapes can satisfy these constraints? The vanishing thickness limit $(t \rightarrow 0)$ corresponds mathematically to enforcing distance preservation exactly, while the bending term selects among possible shapes that satisfy these distance constraints.
\cite{Efrati2009}

The physical system described above leads naturally to a mathematical framework for shape reconstruction. In the simplest case, we seek a mapping $\mathbf{X}(r,s)$ from coordinates on a reference surface to positions in three-dimensional space that preserves a prescribed set of distances.

These distance constraints are expressed through a metric tensor. In its most general form, the squared distance between nearby points is given by
\begin{equation}
  d\ell^2 = \bar{g}_{r r} dr^2 + 2 \bar{g}_{r s} dr ds + \bar{g}_{s s} ds^2.\label{eq-ds}
\end{equation}
This expression can be significantly simplified by choosing special coordinates. Just as a careful choice of coordinates can simplify physical problems, we can choose geodesic coordinates where
\begin{equation}
  d\ell^2 = dr^2 + {\rho(r,s)}^2 ds^2,\label{eq-ds2}
\end{equation}
reducing our three metric components to a single function $\rho(r,s)$. This simplification is always possible locally for any surface in three dimensions, though it may not extend globally.

\begin{figure}
    \centering
    \includegraphics[width=0.45 \textwidth]{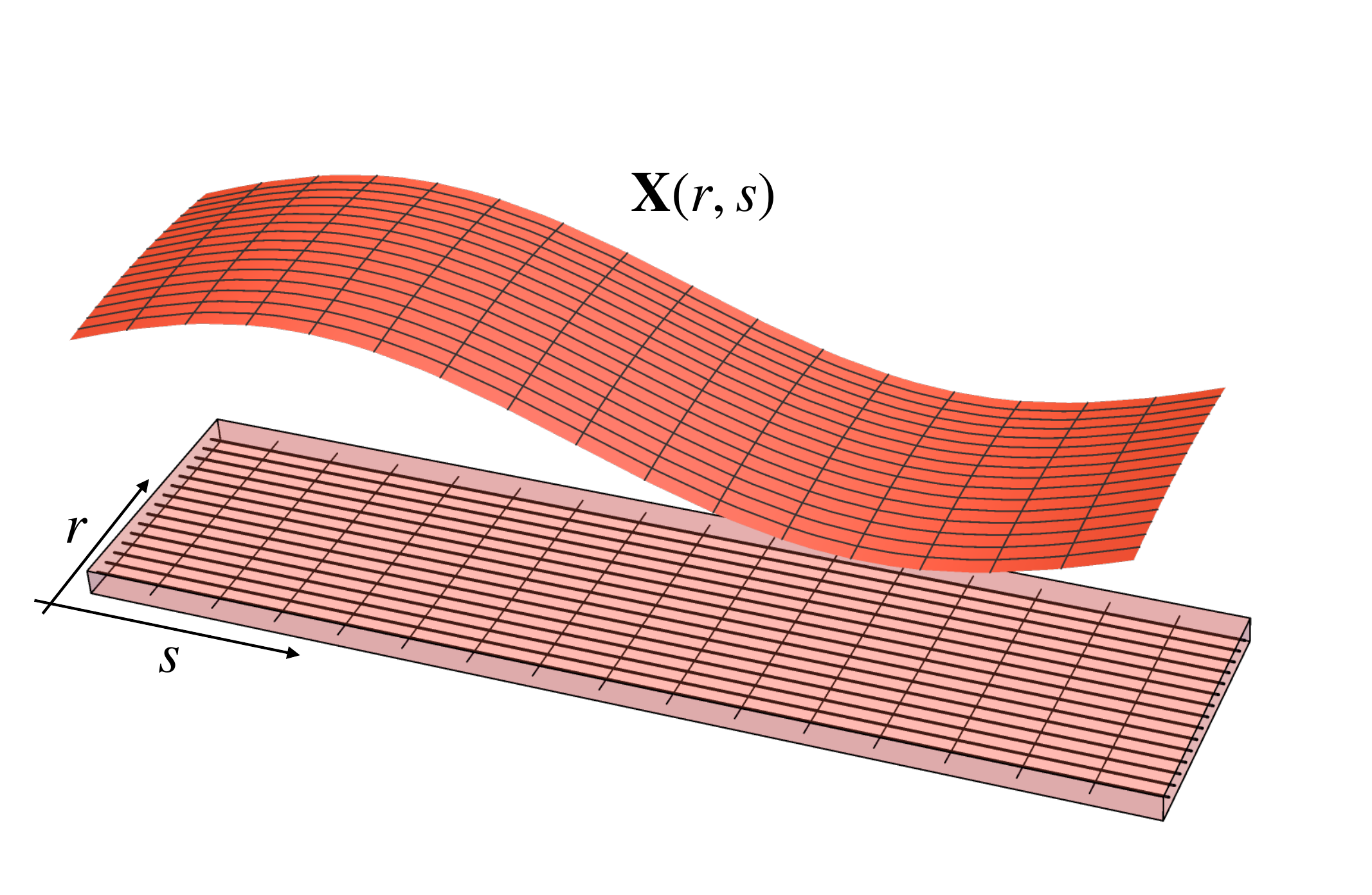}
    \caption{The coordinate system $(r,s)$ defined on a thin, flat strip and its isometric embedding with a prescribed metric $\bar{g}_{ij}(r,s)$.}
    \label{fig:isometric}
\end{figure}

As illustrated in Fig.~\ref{fig:isometric}, this coordinate system provides a natural way to describe the deformation of a thin strip. The reference configuration is parameterized by coordinates $(r,s)$, and the mapping $\mathbf{X}(r,s)$ takes this flat strip into a curved surface while preserving the prescribed metric $\bar{g}_{ij}(r,s)$. The red surface shows one possible embedding that satisfies these metric constraints, though others may exist. 

In these coordinates, shape reconstruction reduces to finding $\mathbf{X}(r,s)$ satisfying three conditions:
\begin{align}
    \left[\partial_r \mathbf{X}(r,s) \right]^2 &= 1, \label{eq:iso1}\\
    \partial_r \mathbf{X}(r,s) \cdot \partial_s \mathbf{X}(r,s) &= 0, \label{eq:iso2}\\
    \left[\partial_s \mathbf{X}(r,s)\right]^2 &= \rho^2(r,s). \label{eq:iso3}
\end{align}
These equations have clear geometric meaning: Equation (\ref{eq:iso1}) shows that the $r$-coordinate measures actual distance; Equation (\ref{eq:iso2}) indicates that the coordinate lines are perpendicular; and Equation (\ref{eq:iso3}) demonstrates that the spacing between $s$-coordinate lines is controlled by  $\rho(r,s)$.

However, these equations alone do not uniquely determine the shape. Just as a physical sheet can bend without stretching, multiple shapes can satisfy these distance constraints (such as flat sheet can be bent to cyliner and cone). There are some special cases when the surface can be uniquely identified using only the intrinsic metric tensor:
1. If boundary conditions provide enough information to estimate the second fundamental form (which captures extrinsic curvature information not contained in the metric tensor; see Example \ref{cont-SR}). 2. When the metric tensor has a unique realization in Euclidean space (e.g., spaces of constant positive curvature which, by the Liebmann theorem \cite{Liebmann1900,doCarmo1976}, must be spheres). 3. When additional physical constraints (such as minimizing elastic energy or imposing symmetry conditions) select a unique configuration among mathematically valid solutions.

However, generally, uniqueness is not guaranteed by the metric alone, as demonstrated by the Gauss-Codazzi(Mainardi) equations \cite{doCarmo1976,Kaneda1990,Kreyszig1991} which show that multiple valid second fundamental forms may satisfy the compatibility conditions for a given metric. This mathematical indeterminacy mirrors the physical reality of thin sheets, which can take multiple shapes while maintaining the same internal distances.

To determine when unique reconstruction is possible, we must move beyond local metric considerations and examine both the local evolution of surfaces and their global topological properties. This dual perspective-connecting local differential equations with global geometric constraints-provides a complete framework for understanding when distance preservation yields unique shapes.

\subsection{Local and Global Properties of Solutions}
Understanding when shape reconstruction has unique solutions requires examining both local and global behavior. This analysis reveals a fundamental distinction between local distance preservation and global shape determination.

\subsubsection{Local Evolution}
In geodesic coordinates, Gauss' \textit{theorema egregium} establishes that the Gaussian curvature depends solely on the first fundamental form coefficients (determined by $\rho$ representing intrinsic shape) and their derivatives, independent of the extrinsic shape (described by the second fundamental form $h_{ij}$) \cite{Kreyszig1991}. Specifically, by Gauss-Codazzi(Mainardi) equations \cite{doCarmo1976,Kaneda1990}, the Gaussian curvature $K$ satisfies
\begin{equation}
    \rho K = h_{rr} h_{ss} - h_{rs}^2,
\end{equation}
where $h_{ij} = \hat{\mathbf{N}} \cdot \partial_i \partial_j \mathbf{X}$ measures how the surface bends in each direction. While this equation relates intrinsic and extrinsic properties, it doesn't uniquely determine the second fundamental form from the metric alone. The Codazzi equations provide additional constraints on the derivatives of  $h_{ij} $, leading to a system of evolution equations that govern how the surface geometry develops from initial conditions:
\begin{equation}\label{eq:evolution}
    \partial_r^2 \mathbf{X} = \left( \frac{-\partial_r^2 \rho + (\hat{\mathbf{N}} \cdot \partial_r \partial_s \mathbf{X})^2 }{\hat{\mathbf{N}} \cdot \partial_s^2 \mathbf{X} } \right) \hat{\mathbf{N}}.
\end{equation}

Given an initial curve at $r=0$ satisfying
\begin{equation}\label{eq:continuumembed}
    \partial_s^2 \mathbf{X} = - \rho \partial_r \rho \partial_r \mathbf{X} + \frac{\partial_s \rho}{\rho} \partial_s \mathbf{X} + f(s) \hat{\mathbf{N}},
\end{equation}

We can evolve the surface uniquely (up to rigid motions) in a neighborhood where the geodesic coordinates remain well-defined, $h_{ss}(r,s)$ stays non-zero, and $\rho(r,s)$ remains analytic.

\subsubsection{Global Behavior}
While local evolution is unique under suitable conditions, global behavior is more subtle. The shape of the surface depends crucially on the function $f(s)$ in equation (\ref{eq:continuumembed}), which determines how the surface bends initially. Different choices of $f(s)$ can lead to different global shapes.

This behavior parallels the physics of thin sheets: while the metric determines local distances (like the no-stretch condition), the actual shape depends on how the surface bends (like the bending energy term). To quantify this, consider two points $P_1(r_1,s_1)$ and $P_2(r_2,s_2)$ on the surface. We measure:
\begin{itemize}
    \item Local metric distance:
    \begin{equation}
        d\ell^2 = dr^2 + \rho(r,s)^2ds^2
    \end{equation}
    \item Euclidean distance (through space):
    \begin{equation}
        d_E = ||\mathbf{X}(r_1,s_1) - \mathbf{X}(r_2,s_2)||
    \end{equation}
\end{itemize}

The geodesic distance is determined entirely by the metric (through $\rho$), but the Euclidean distance depends on the global shape. This leads to the non-uniqueness of shape reconstruction from distance constraints alone. The following example illustrates these differences by showing two fundamentally different surfaces that satisfy the same intrinsic metric but differ in their embedding in Euclidean space:
\begin{example}
Consider a surface with constant Gaussian curvature $K = 0$ parametrized in geodesic coordinates with $\rho(r,s) = 1$ (constant). At $r=0$, let's specify the initial curve with two different choices of $f(s)$: For the first embedding $\mathbf{X}_1$, choose $f_1(s) = 0$ (plane), and for the second embedding $\mathbf{X}_2$, choose $f_2(s) = 1$ (cylinder). This choice satisfies our evolution equation because $\partial_r \rho = 0$ and $\partial_r^2 \rho = 0$ (since $\rho$ is constant), $\partial_s \rho = 0$ (again, due to constant $\rho$). From equation (\ref{eq:continuumembed}), this simplifies to: $\partial_s^2 \mathbf{X} = f(s) \hat{\mathbf{N}}$.

For the plane embedding with $f_1(s) = 0$, the equation becomes $\partial_s^2 \mathbf{X}_1 = 0$, which gives us a parametrization $\mathbf{X}_1(r,s) = (r, s, 0)$.

For the cylinder embedding with $f_2(s) = 1$, the equation becomes $\partial_s^2 \mathbf{X}_2 = \hat{\mathbf{N}}$. This corresponds to a constant normal curvature in the $s$-direction, resulting in a cylindrical surface. Solving this equation yields $\mathbf{X}_2(r,s) = (r, \sin(s), 1-\cos(s))$, where we've chosen the normal vector consistently with our coordinate system.

Both surfaces have the same first fundamental form ($d\ell^2 = dr^2 + ds^2$), ensuring identical geodesic distances. However, the Euclidean distances differ. For points $P_1 = (0,0)$ and $P_2 = (0,\pi)$, the geodesic distance is $d_G(P_1,P_2) = \pi$ on both surfaces, but the Euclidean distances are:

For the plane: $\|\mathbf{X}_1(P_1) - \mathbf{X}_1(P_2)\| = \|(0,0,0) - (0,\pi,0)\| = \pi$

For the cylinder: $\|\mathbf{X}_2(P_1) - \mathbf{X}_2(P_2)\| = \|(0,0,0) - (0,0,2)\| = 2$

This example demonstrates that surfaces with identical intrinsic geometry (same metric tensor) can have different Euclidean distances when embedded in 3D space according to different second fundamental forms.
\end{example}

This distinction between preserved geodesic distances and varying Euclidean distances has profound implications. A striking example is the flat torus: while it cannot be smoothly ($C^\infty$: infinitely differentiable) isometrically embedded in $\mathbb{R}^3$ \cite{Hilbert1901, Nash1954}, Borrelli et al. \cite{FlatToriEmb} showed that a $C^1$ (continuously differentiable once) isometric embedding is possible through a fractal-like structure of wrinkles. This demonstrates how different regularity conditions ($C^\infty$ vs $C^1$) affect the existence of isometric embeddings while preserving the intrinsic geometry.

\subsection{From Graph Embeddings to Physical Realizations}
Having established the theoretical foundations for continuous surfaces, we now turn to their discrete counterparts. The fundamental distinction between distance preservation and shape determination that we observed in continuous systems manifests differently in discrete structures, but with similar ramifications. In discrete systems, this distinction appears through three interrelated problems: graph isometric embedding, the graph realization problem, and physical shape reconstruction.

\begin{definition}[Graph Isometric Embedding by Graham and Winkler \cite{Graham1985}]
For a finite, connected, undirected graph $G = (V, E)$, the graph metric $d_G: V \times V \to \mathbb{N}$ assigns to each pair of vertices $x,y \in V$ the number of edges in the shortest path between them. An embedding $\lambda: V \to M$ into a metric space $(M, d_M)$ is isometric if
\begin{equation}
    d_M(\lambda(x), \lambda(y)) = d_G(x, y)
\end{equation}
for all $x, y \in V$.
\end{definition}
\begin{definition}[Graph Embedding Problem (Graph Realization)]
For a finite connected undirected graph $G = (V, E)$ with a partial distance matrix $D$, where some entries $d_{ij}$ correspond to prescribed edge lengths between vertices $(i,j) \in E$, the graph realization problem seeks a mapping $\phi: V \to \mathbb{R}^d$ such that:
\begin{equation}
    \|\phi(i) - \phi(j)\| = d_{ij} \quad \text{for all } (i,j) \in E \text{ where } d_{ij} \text{ is defined}
\end{equation}
Here, $D$ may be incomplete (containing only some pairwise distances) or complete (containing all pairwise distances).
\end{definition}

The number of possible realizations is fundamentally connected to the rigidity of the graph. For non-rigid (flexible) graphs (technically, we refer to infinitesimally rigid graphs), continuous deformations are possible while maintaining all edge lengths, leading to infinitely many realizations which correspond to a solution space of dimension at least one, rather than discrete points. A graph is considered rigid when no infinitesimal deformation is possible that preserves all edge lengths to first order, and a minimally rigid graph is one that becomes flexible if any edge is removed-in other words, it has exactly the edges needed for rigidity, with no redundant constraints. By definition, flexible graphs cannot have a finite number of discrete realizations, and for minimally rigid graphs, this number is typically bounded by a function of the number of vertices. The finiteness of solutions can be determined by analyzing the rank of the rigidity matrix \cite{Hendrickson1992,Gortler2010} and the completeness of the distance matrix \cite{Laurent1998,Alfakih1999}. While higher-order rigidity conditions exist and are relevant to phenomena such as bistable configurations and buckling \cite{Connelly1996, Chi2022} (continuum \cite{Kebadze2004}), since these studies focus on transitions between solutions (equilibrium points) rather than the solution space itself, we restrict our attention to first-order rigidity.

A graph isometric embedding preserves all pairwise path distances, making it a strong condition that may not always be realizable in a given dimension. (We note that some literature defines isometric embedding as implying equal edge lengths; however, we follow the definition from Graham and Winkler where distances are preserved between all pairs of vertices.) This relationship between complete path distances and realizability directly connects to the rank of the distance matrix: while edge lengths alone provide information analogous to the first fundamental form in continuous surfaces, additional distances between non-adjacent vertices constrain the dihedral angles, similar to how the second fundamental form determines the shape in the continuous case. This distinction parallels Gauss's theorema egregium and Gauss-Codazzi(Mainardi) in differential geometry, which establishes that Gaussian curvature depends solely on the intrinsic metric, while the complete shape determination requires additional information about mean curvature.

In contrast to isometric embedding, shape reconstruction and graph realization problems typically focus on preserving only specified edge lengths, which generally leads to multiple possible solutions. This approach varies across different problem domains: graph isometric embedding preserves all pairwise path distances, while the graph embedding problem maintains only specified edge lengths. Shape reconstruction extends this by combining geometric and physical constraints. Physical shape reconstruction thus encompasses two fundamental aspects that work in tandem: finding a configuration that satisfies (distance) constraints and determining all possible realizations of this configuration. This relationship can be summarized as shown in Table \ref{table1}.
\begin{table*}
\begin{center}
\begin{tabular}{|l||l|l|l|}
\hline
Problem & \multicolumn{1}{l||}{Mapping} & \multicolumn{1}{l||}{Constraints} & Uniqueness \\
\hline\hline
Isometric Graph & \multirow{2}{*}{$G \to \mathbb{R}^d$} & All pairwise path distances & Realizable or \\
Embedding & & (complete distance matrix $D$) & not realizable \\
\hline
Graph Embedding Problem & \multirow{2}{*}{$G \to \mathbb{R}^d$} & Specific edge lengths & \multirow{2}{*}{Multiple solutions} \\
(Realization) & & (complete/incomplete $D$) & \\
\hline
Shape & $\mathbb{R}^d \to$ & \multirow{2}{*}{Both types} & Inherits \\
Reconstruction & $\mathbb{R}^d$ & & multiplicity \\
\hline
\end{tabular}
\end{center}
\caption{The relationship between isometric embedding, graph embedding problem, and shape reconstruction.}
\label{table1}
\end{table*}

When implementing numerical solutions for continuous elasticity problems, the theoretical insights developed above have direct implications for computational practice. The choice of discretization grid becomes crucial, representing a critical bridge between continuous theory and discrete computation. While square or triangular grids are commonly used for their simplicity, they can introduce significant challenges that fundamentally alter the mathematical properties of the system being modeled. Efrati et al. \cite{Efrati2009} demonstrated that elastic bodies with non-Euclidean intrinsic geometry can undergo transitions between stretching-dominated and bending-dominated configurations as parameters like thickness vary. This reveals how physical discretization parameters can fundamentally alter system behavior. 

This discrepancy arises from two main sources: First, the discrete nature of the grid only preserves local properties (neighboring distances), which can lead to unexpected embeddings even when the structure appears to be $C^1$ continuous, as demonstrated by Marder and Papanicolaou \cite{Marder2006} in their work on imposing metrics on surfaces. Unlike continuous systems where infinitesimal changes preserve smoothness (i.e., $d f(x)/d x\approx d(f(x+\delta))/d(x+ \delta)$ for sufficiently small $\delta$), discrete systems lack this natural continuity due to the triangle inequality, resulting in edges that behave independently without inherent relationships between neighboring vertices beyond explicit edge constraints (opposite case - graph embedding to Riemannian embedding is discussed in \cite{ren2018}). These independent behaviors can introduce directions of deformation not present in the continuous system \cite{Grinspun2008}. Second, as shown by Connelly \cite{Connelly1979}, not all triangulated surfaces are rigid even though their continuous counterparts homeomorphic to a sphere have unique realizations. 

This fundamental difference between discrete and continuous systems in terms of rigidity manifests in several significant ways. The reconstructed shape from discrete property can not satisfy all properties compared to the continuous system \cite{Wardetzky2007,Marder2006}, while energy minimization processes may yield multiple solutions \cite{Seung1988}. Moreover, unlike continuous smooth surface, in the discrete case, adding boundary conditions as continuous case may not be helpful much since dihedral angles are not periodic or constant in general. Additionally, certain configurations may satisfy discrete constraints yet violate continuous smoothness principles \cite{Wallner2011,Marder2006}. These issues underscore the essential nature of careful grid structure consideration for reliable shape reconstruction. Hotz et al. \cite{Hotz2004} also pointed out these differences and developed method using curvature estimating tools to check. Our approach addresses these challenges through a comprehensive strategy that implements appropriate rigidity conditions for discrete structures, maintains controlled relationships between (distance) constraints, and ensures compatibility between local and global geometric properties.

\begin{claim}\label{claim1}
For surfaces that admit multiple realizations, reconstructions with insufficient constraints on dihedral angles (which determine the second fundamental form) tend to exhibit greater local fluctuations than the desired shape due to the wrong choice of the solution compared to true solution.
\end{claim}

One solution to these issues is to add sufficient distance constraints (or sufficient additional constraints or more boundary values) to ensure unique realization in cases where the continuum shape is expected to be unique. This approach bridges the gap between discrete and continuous systems by enforcing both local and global geometric constraints.

This claim, which we will rigorously demonstrate through simulations in Section~\ref{sec:simulation}, provides a practical diagnostic tool for shape reconstruction problems: local fluctuations in reconstructed shapes serve as quantifiable indicators of insufficient geometric constraints. This insight directly connects our theoretical analysis to computational implementation, offering a path to more reliable discrete approximations of continuous systems.

To address these challenges systematically, we propose adding sufficient distance constraints to ensure unique realization in cases where the continuum shape is expected to be unique. This approach bridges the gap between discrete and continuous systems by enforcing both local and global geometric constraints. In the following section, we demonstrate how this approach leads to significant improvements in shape reconstruction accuracy and stability across diverse geometric configurations.
\section{Results}
\subsection{Counting Methods for Graph Realizations}
Shape reconstruction inherits solution multiplicity from the graph embedding problem, necessitating an understanding of possible realizations. While rigidity theory provides bounds for minimally rigid graphs (such as convex triangulated surfaces, see Fig.~\ref{fig:triangulation}), practical applications often involve more general structures. Moreover, as discussed in the previous section, even triangulated surfaces can exhibit unexpected behaviors due to the discrete nature of the constraints.

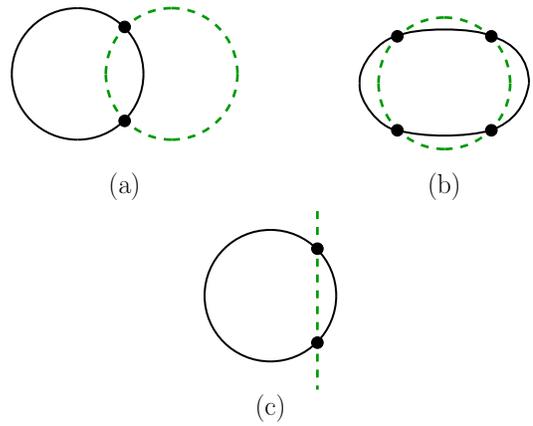
\begin{figure}[H] 
		\centering
\begin{tikzpicture}[scale=0.50, transform shape]
	\begin{pgfonlayer}{nodelayer}
		\node [style=none] (0) at (-7.5, 0.75) {};
		\node [style=none] (1) at (-4, 0.75) {};
		\node [style=none] (2) at (-5.75, 2.5) {};
		\node [style=none] (3) at (-5.75, -1) {};
		\node [style=none] (4) at (-5, 0.75) {};
		\node [style=none] (5) at (-1.5, 0.75) {};
		\node [style=none] (6) at (-3.25, 2.5) {};
		\node [style=none] (7) at (-3.25, -1) {};
		\node [style=none] (8) at (1.75, 0.5) {};
		\node [style=none] (9) at (6.25, 0.5) {};
		\node [style=none] (10) at (2.75, 1.75) {};
		\node [style=none] (11) at (2.75, -0.75) {};
		\node [style=none] (12) at (2.25, 0.5) {};
		\node [style=none] (13) at (5.75, 0.5) {};
		\node [style=none] (14) at (4, 2.25) {};
		\node [style=none] (15) at (4, -1.25) {};
		\node [style=none] (16) at (5.25, 1.75) {};
		\node [style=none] (17) at (5.25, -0.75) {};
		\node [style=blackdot] (18) at (-4.5, 2) {};
		\node [style=blackdot] (19) at (-4.5, -0.5) {};
		\node [style=blackdot] (20) at (2.75, 1.75) {};
		\node [style=blackdot] (21) at (5.25, 1.75) {};
		\node [style=blackdot] (22) at (2.75, -0.75) {};
		\node [style=blackdot] (23) at (5.25, -0.75) {};
		\node [style=none] (24) at (-4.5, -2.25) {\huge (a)};
		\node [style=none] (25) at (4, -2.25) {\huge (b)};
	\end{pgfonlayer}
	\begin{pgfonlayer}{edgelayer}
		\draw [style=blackedge,bend left=45] (0.center) to (2.center);
		\draw [style=blackedge,bend left=45] (2.center) to (1.center);
		\draw [style=blackedge,bend left=45] (1.center) to (3.center);
		\draw [style=blackedge,bend left=45] (3.center) to (0.center);
		\draw [style=greenedge, bend left=45] (4.center) to (6.center);
		\draw [style=greenedge, bend left=45] (6.center) to (5.center);
		\draw [style=greenedge, bend left=45] (5.center) to (7.center);
		\draw [style=greenedge, bend left=45] (7.center) to (4.center);
		\draw [style=blackedge,bend left=45, looseness=0.75] (11.center) to (8.center);
		\draw [style=greenedge, bend left=45] (12.center) to (14.center);
		\draw [style=greenedge, bend left=45] (14.center) to (13.center);
		\draw [style=greenedge, bend left=45] (13.center) to (15.center);
		\draw [style=greenedge, bend left=45] (15.center) to (12.center);
		\draw [style=blackedge,bend left, looseness=0.50] (10.center) to (16.center);
		\draw [style=blackedge,in=90, out=-15] (16.center) to (9.center);
		\draw [style=blackedge,bend left] (9.center) to (17.center);
		\draw [style=blackedge,bend left=15, looseness=0.75] (17.center) to (11.center);
		\draw [style=blackedge,bend left=45, looseness=0.75] (8.center) to (10.center);
	\end{pgfonlayer}
\end{tikzpicture}
\begin{tikzpicture}[scale=0.50, transform shape]
	\begin{pgfonlayer}{nodelayer}
		\node [style=none] (20) at (1.25, 0.75) {};
		\node [style=none] (21) at (4.75, 0.75) {};
		\node [style=none] (22) at (3, 2.5) {};
		\node [style=none] (23) at (3, -1) {};
		\node [style=none] (26) at (4.25, 3) {};
		\node [style=none] (27) at (4.25, -1.75) {};
		\node [style=blackdot] (28) at (4.25, 2) {};
		\node [style=blackdot] (29) at (4.25, -0.5) {};
		\node [style=none] (31) at (3, -2.25) {\huge (c)};
	\end{pgfonlayer}
	\begin{pgfonlayer}{edgelayer}
		\draw [style=blackedge,bend left=45] (20.center) to (22.center);
		\draw [style=blackedge,bend left=45] (22.center) to (21.center);
		\draw [style=blackedge,bend left=45] (21.center) to (23.center);
		\draw [style=blackedge,bend left=45] (23.center) to (20.center);
		\draw [style=greenedge] (26.center) to (27.center);
	\end{pgfonlayer}
\end{tikzpicture}
\caption{(a) Intersection points of two circles. (b) Intersection points of a ellipse and a circle. The maximum number of intersection points for any two second-order curves is four, as per Bézout's Theorem. (c) The same intersection points as in (a), but using a line and a circle.
}
\label{fig:linearity-circles}
\end{figure}

In practical applications, we often work with graphs that are not minimally rigid, making it difficult to estimate the number of possible realizations directly from rigidity theory. Therefore, we develop counting methods applicable to a broader class of graphs. These methods not only help estimate the number of possible embeddings but also guide the addition of constraints needed to achieve a desired unique shape. Our approach is based on analyzing the intersection points of a circle using B\'ezout's Theorem.
\paragraph{Algebraic Approach Using B\'ezout's Theorem}
Our counting method relies on B'ezout's Theorem, which states that if $A$ and $B$ are polynomial equations of degrees $m$ and $n$ respectively, with no nonconstant common factor, their intersection contains exactly $mn$ points, counting multiplicities \cite{Hilmar10}. This provides a powerful tool for analyzing intersection configurations, as demonstrated in our figures: two circles can have a maximum of 2 intersection points (Fig.\ref{fig:linearity-circles}(a)), while two second-order curves can intersect at up to 4 points (Fig.\ref{fig:linearity-circles}(b)).

 By leveraging this result, we can systematically compute potential embedding configurations for graphs. For instance, two overlapping circles can have a maximum of 2 intersection points, as shown in Fig. \ref{fig:linearity-circles} (a). However, for two second-order polynomials, the maximum number of intersection points will be 4, according to B\'ezout's Theorem, as illustrated in Fig. \ref{fig:linearity-circles} (b).

\paragraph{Dimensional Reduction of Sphere Equations}
To make this approach practical, we show how $n$-dimensional sphere equations can be reduced to $(n-1)$-dimensional problems through linear algebra. This reduction significantly simplifies the computation of intersection points.

Consider three \((d-1)\)-dimensional sphere equations for a point \(x_i = (x_i^1, x_i^2, \dots, x_i^d) \in \mathbb{R}^d\). Suppose the sphere centers are at \(p_1, p_2,\) and \(p_3\), where \(p_i = (p_i^1, p_i^2, \dots, p_i^d)\) with radii \(d_{1i}, d_{2i},\) and \(d_{3i}\), respectively. The corresponding sphere equations are given by:

\begin{flalign*} 
&(x_i^1 - p_1^1)^2 + (x_i^2 - p_1^2)^2 + \dots + (x_i^d - p_1^d)^2 = d_{1i}^2 \\ 
&(x_i^1 - p_2^1)^2 + (x_i^2 - p_2^2)^2 + \dots + (x_i^d - p_2^d)^2 = d_{2i}^2 \\ 
&(x_i^1 - p_3^1)^2 + (x_i^2 - p_3^2)^2 + \dots + (x_i^d - p_3^d)^2 = d_{3i}^2 
\end{flalign*}

By subtracting these equations, we derive:

\begin{flalign*} 
&2\sum_{j=1}^d x_i^j (-p_1^j + p_2^j) + \sum_{j=1}^d \left[(p_1^j)^2 - (p_2^j)^2\right] =d_{1i}^2 - d_{2i}^2 \\ 
&2 \sum_{j=1}^d x_i^j (-p_1^j + p_3^j) + \sum_{j=1}^d \left[(p_1^j)^2 - (p_3^j)^2\right]= d_{1i}^2 - d_{3i}^2 
\end{flalign*}

If the vectors \((-p_1^j + p_2^j)\) and \((-p_1^j + p_3^j)\) are linearly independent (i.e., the rows of the matrix \(\begin{bmatrix} p_1^j - p_2^j \\ p_1^j - p_3^j \end{bmatrix}\) are linearly independent or have different slopes), then we can eliminate two variables from \((x_i^1, \dots, x_i^d)\). Consequently, the solution space for the \(n\)-dimensional sphere equations can be reduced to lower-dimensional sphere equations. This reduction results from the linearity of intersecting or overlapping spheres, as illustrated in Fig. \ref{fig:linearity-circles} (c).

\paragraph{From Circles to Lines}
The dimensional reduction principle starts with a simple case. The intersection points of two circle equations can be transformed into simpler geometric objects:
\begin{flalign*} 
&(x - p_{1x})^2 + (y - p_{1y})^2 = d_{1}^2 \\ 
&(x - p_{2x})^2 + (y - p_{2y})^2 = d_{2}^2 
\end{flalign*}
Subtracting these equations yields:
\begin{flalign} 
&(x - p_{1x})^2 + (y - p_{1y})^2 = d_{1}^2 \notag \\ 
&2x(p_{1x} - p_{2x}) + 2y(p_{1y} - p_{2y}) \label{eq:two-circles1} \\
&+ (p_{2x}^2 - p_{1x}^2 + p_{2y}^2 - p_{1y}^2) = d_{2}^2 - d_{1}^2\label{eq:two-circles2} 
\end{flalign}
For fixed centers and radii, equation (\ref{eq:two-circles2}) reduces to a linear equation $ax + by = c$, demonstrating how quadratic constraints can be partially linearized.

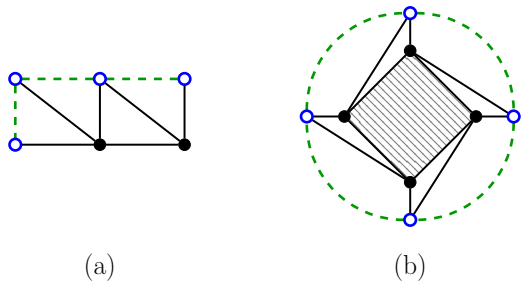
\begin{figure}[H]
    \centering
\begin{tikzpicture}[scale=0.50, transform shape]
	\begin{pgfonlayer}{nodelayer}
		\node [style=bluedot] (0) at (-8.5, 3.25) {};
		\node [style=bluedot] (1) at (-8.5, 1.5) {};
		\node [style=blackdot] (2) at (-4, 1.5) {};
		\node [style=bluedot] (3) at (-4, 3.25) {};
		\node [style=blackdot] (4) at (-6.25, 1.5) {};
		\node [style=bluedot] (5) at (-6.25, 3.25) {};
		\node [style=blackdot] (6) at (0.25, 2.25) {};
		\node [style=blackdot] (7) at (3.75, 2.25) {};
		\node [style=blackdot] (8) at (2, 4) {};
		\node [style=blackdot] (9) at (2, 0.5) {};
		\node [style=bluedot] (10) at (-0.75, 2.25) {};
		\node [style=bluedot] (11) at (4.75, 2.25) {};
		\node [style=bluedot] (12) at (2, 5) {};
		\node [style=bluedot] (13) at (2, -0.5) {};
		\node [style=none] (14) at (-6.25, -1.75) {\huge (a)};
		\node [style=none] (15) at (2, -1.75) {\huge (b)};
	\end{pgfonlayer}
	\begin{pgfonlayer}{edgelayer}
		\fill[gray, thick, pattern=north west lines, pattern color=gray] (6.center) to (8.center) to (7.center) to (9.center) to cycle;
		\draw [style=greenedge] (0) to (1);
		\draw [style=blackedge](1) to (2);
		\draw [style=blackedge](3) to (2);
		\draw [style=blackedge](0) to (4);
		\draw [style=greenedge] (0) to (5);
		\draw [style=greenedge] (5) to (3);
		\draw [style=blackedge](5) to (4);
		\draw [style=blackedge](5) to (2);
		\draw [style=blackedge](6) to (8);
		\draw [style=blackedge](8) to (7);
		\draw [style=blackedge](7) to (9);
		\draw [style=blackedge](9) to (6);
		\draw [style=greenedge, bend left=45] (10) to (12);
		\draw [style=greenedge, bend left=45] (12) to (11);
		\draw [style=greenedge, bend left=45] (11) to (13);
		\draw [style=greenedge, bend left=45] (13) to (10);
		\draw [style=blackedge](12) to (8);
		\draw [style=blackedge](10) to (6);
		\draw [style=blackedge](9) to (13);
		\draw [style=blackedge](7) to (11);
		\draw [style=blackedge](12) to (6);
		\draw [style=blackedge](10) to (9);
		\draw [style=blackedge](13) to (7);
		\draw [style=blackedge](11) to (8);
	\end{pgfonlayer}
\end{tikzpicture}
    \caption{An illustration of strip and annulus (genus is highlighted as gray pattern). The discrete strip is constructed with two $I_3$ (left), and the annulus is composed of four $I_3$ (right). Boundary lines are represented as dashed green lines, and pinned points are white filled blue dots. Unpinned points and edges are presented as black dots and lines.}
    \label{fig:strip-annulus}
\end{figure}

\paragraph{Applications to Geometric Structures}
The dimensional reduction principle enables systematic counting of solutions. When we have $(d+1)$ equations of $(d-1)$-spheres, we obtain $d$ linear equations in $d$ coordinates, leading to unique solutions. This provides a foundation for counting realizations in different geometric settings: Discrete Strip (genus 0): Solutions countable from one or both ends (Fig. \ref{fig:strip-annulus} (a)), and Discrete Annulus (genus 1): Direction-independent solution counting (Fig. \ref{fig:strip-annulus} (b)). These counting methods connect directly to the $I_d$ structures discussed in the next section, where each vertex has $d$ linearly independent edges.

These geometric analyses lead to specific bounds on the number of realizations based on graph structure:

\begin{theorem}[Realization Bounds]
For a graph constructed in $\mathbb{R}^d$:
\begin{itemize} 
\item With $n$ copies of $I_d$ (and pinned vertices):
    \begin{itemize}
        \item Strip structure: $N_G \leq 2^n$
        \item Annulus structure: $2^n \leq N_G \leq 4^n$
    \end{itemize}
\item With $(n-1)$ copies of $I_d$ and at least one $I_{d+1}$:
    \begin{itemize}
        \item $N_G \leq 2^n$, with exact count depending on subgraph connections
    \end{itemize}
\end{itemize}
where $N_G$ denotes the number of realizations for graph $G$.
\end{theorem}

To illustrate these bounds, we can examine two key examples: in Fig.\ref{fig:strip-annulus} (a), two $I_3$ structures yield a maximal embedding count of $2^2$, while in Fig.\ref{fig:strip-annulus} (b), direction-independent counting produces bounds of $2^4 \leq N_G \leq 4^4$. These bounds generalize to $n$ vertices. For example, with three pinned vertices forming a triangle, we obtain $2^{n-3} \leq N_G \leq 4^{n-3}$. To compare these bounds with known results:
\begin{definition} 
$M_d(n)$ denotes the maximum number of realizations of minimally rigid graphs with $n$ vertices in $d$-dimensional space. 
\end{definition}

\begin{table}[ht]
	\centering 
	\begin{tabular}{c c c c c c}
		\hline\hline                        
		$|V|=n$ & 6 & 7 & 8 & 9 & 10 \\ 
		\hline                  
		lower ($2^{n-3}$)  & 8 & 16 & 32 & 64 & 128 \\ 
		upper ($4^{n-3}$) & 64 & 256 & 1024 & 4096 & 16384  \\ 
		\hline 
		$M_3(n)$ \cite{Grasegger18}  & 16 & 48 & 160 & 640 & 2560  \\ 
		\hline
	\end{tabular}
	\caption{$M_3(n)$ and upper bounds among all minimally rigid graphs with $|V|$ vertices for $6\leq |V|\leq 10$, as well as lower and upper bounds for pinned discrete annulus for $6\leq |V|\leq 10$ with $|P|=3$.}
	\label{table:3d-embedding}
\end{table}

As shown in Table \ref{table:3d-embedding}, our estimated bounds are consistent with the measured values of $M_3(n)$ from \cite{Grasegger18}, providing practical guidelines for estimating realization counts in physical applications.
\subsection{Construction of Graphs with Finite Realizations}
The existence of finite realizations can be analyzed through multiple complementary approaches. These include algebraic conditions that relate edge constraints to degrees of freedom, sequential computation methods (marching methods) such as trilateration and linear matrix approaches, and Cayley-Menger determinant properties for distance matrix completion.

\subsubsection{Algebraic Conditions for Finite Realizations}
For m unknown vertices in $\mathbb{R}^d$, the relationship between edge constraints $|E|$ and degrees of freedom determines the nature of realizations: Multiple finite realizations occur when $dm \leq |E| < \binom{m+1}{2}$, while a unique realization exists when $\binom{m+1}{2} \leq |E|$. These conditions assume algebraically independent edge lengths.

The Cayley-Menger determinant \cite{Blumenthal1970, Havel2002} provides a fundamental tool for verifying realizability and completing missing distances in the distance matrix. (Definition \ref{CMDef}) For a structure in $\mathbb{R}^3$, realizability requires a series of conditions on substructures: $CM(P_0,...,P_3) = 0$ (for each tetrahedron) and $CM(P_0,...,P_4) = 0$ (for connected tetrahedra) where $CM$ denotes the Cayley-Menger determinant. The first condition ensures local realizability of each tetrahedron, while the second ensures compatibility between connected tetrahedra. A detailed treatment of these conditions and their application to distance matrix completion is provided in the Appendix.

\subsubsection{Sequential Construction Framework}
To develop systematic construction methods, we first introduce key concepts from rigidity theory.

\begin{definition}[$I_n$ Structure]
$I_n$ represents a set of $n$ edges connected to a vertex, characterized by nonzero angles between each pair of edges and nonzero angles between all edges and their adjacent hyperplanes. For example, $I_3$ in $\mathbb{R}^3$ forms a tripod with three edges meeting at non-degenerate angles.
\end{definition}

\begin{definition}[Rigidity Matrix \cite{Trinh2016}]
Consider a graph $G(V,E)$ with $n$ vertices and $m$ edges. Let $H \in \mathbb{R}^{m \times n}$ be the incidence matrix of $G$, and $H^T H$ be the Laplacian matrix of $G$. The rigidity matrix is defined as 
\[
R = \Lambda(p_i - p_j)^T (H \otimes \mathbb{I}_d),
\]
where $\Lambda$ is a diagonal matrix and $\mathbb{I}_d$ denotes the identity matrix of size $d \times d$, corresponding to the dimension of the space. The symmetric rigidity matrix is defined as 
\[
M = R^T R \in \mathbb{R}^{dn \times dn}
\]
for $d$-dimensional space.
\end{definition}

The symmetric rigidity matrix $M$ provides information about graph rigidity through its rank:
\begin{equation}
    \rank M = dn - \binom{d+1}{2} \quad \text{for } d=3
\end{equation}
This rank condition ensures infinitesimal rigidity, while the pattern of zeros in $M$ indicates how close the graph is to being complete.

\begin{figure}
    \centering
    \begin{tabular}{c c}
        \includegraphics[width=0.4\linewidth]{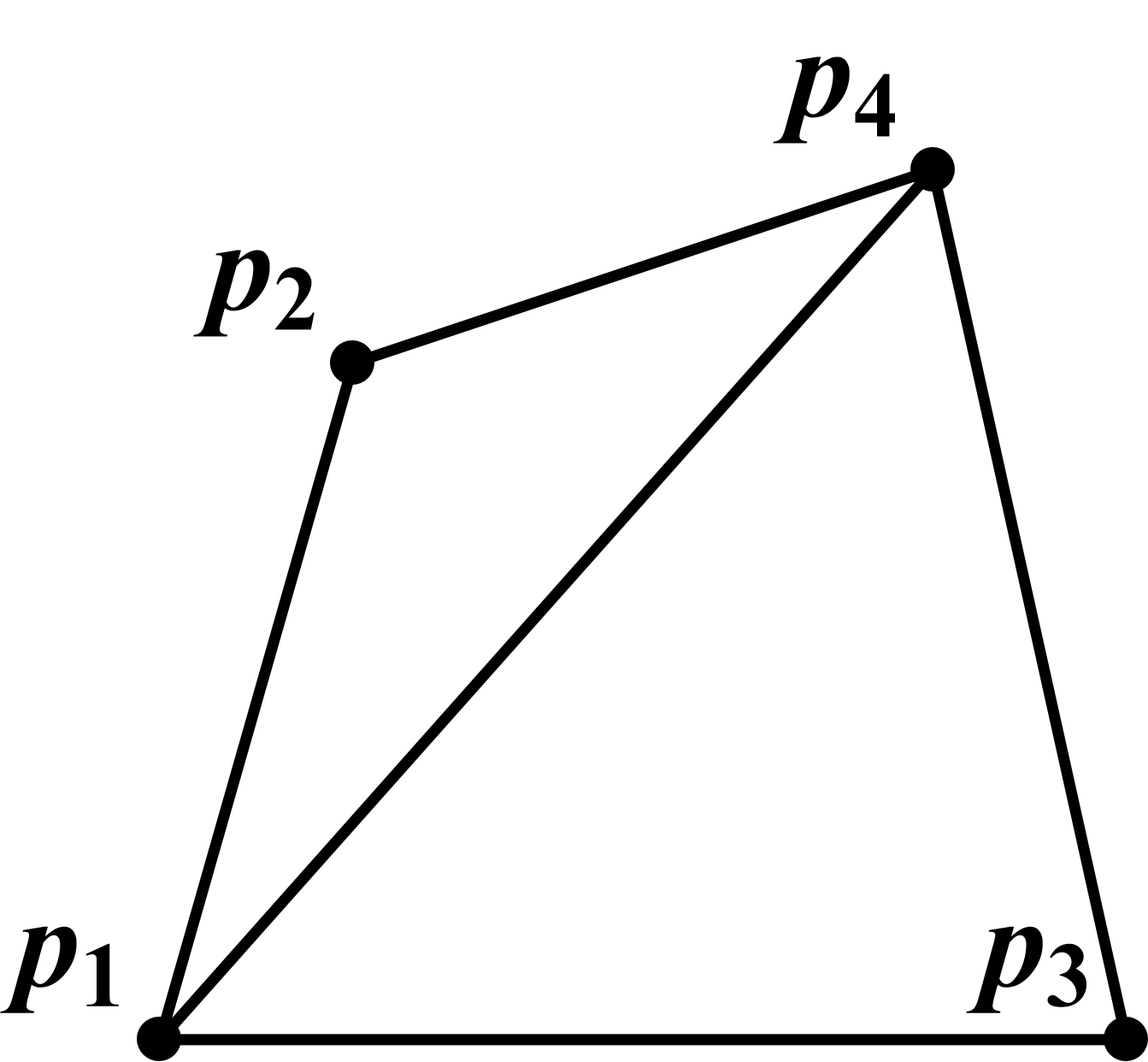}&
        \includegraphics[width=0.4\linewidth]{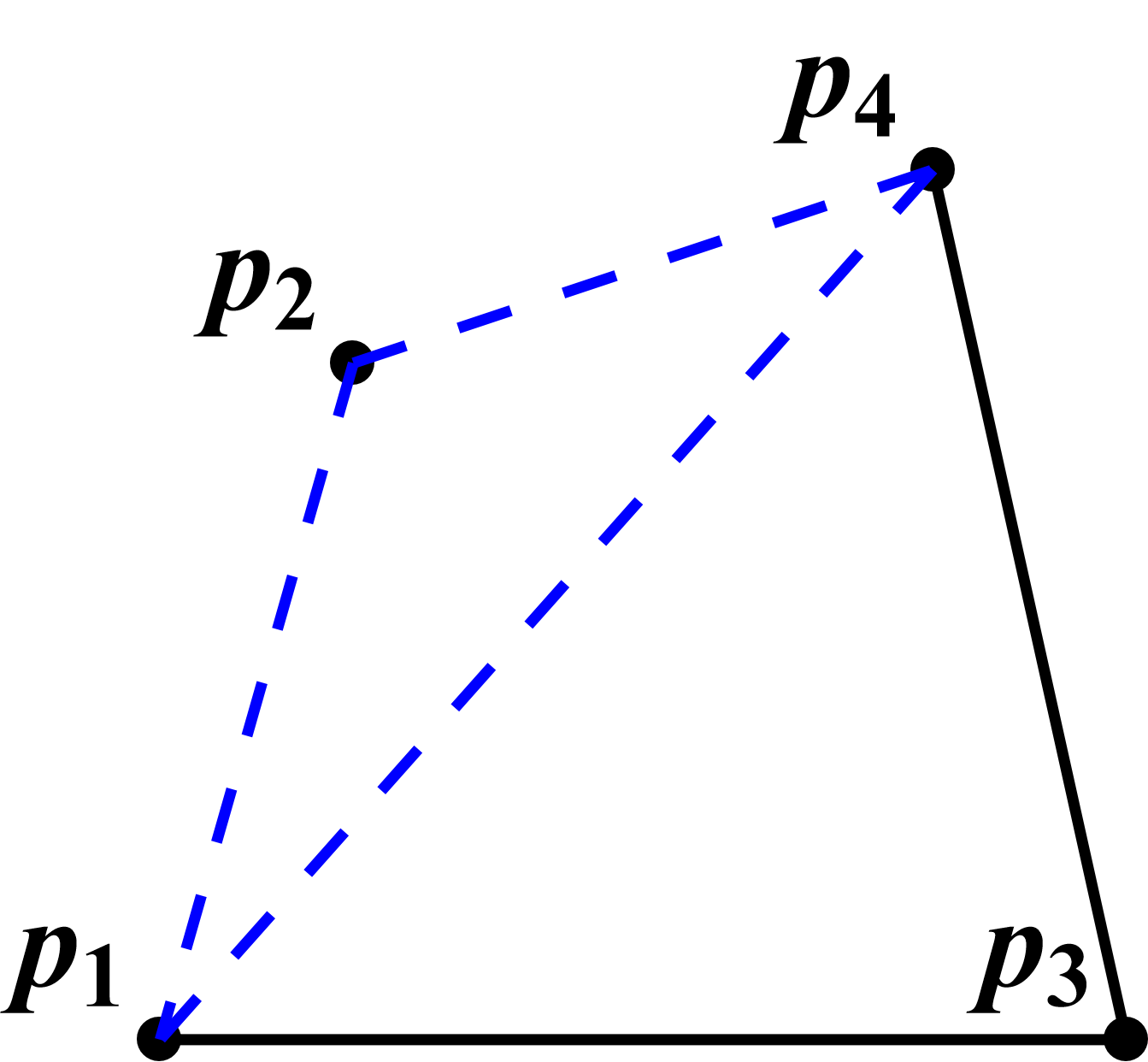}\\
         (a) two neighboring triangles & (b) pinned vertices with $I_2$
    \end{tabular}
    \caption{Caption}
    \label{fig:ex2d}
\end{figure}

This symmetric rigidity matrix has a similar form to the Laplacian matrix. Using simpler notation instead of full coordinate vectors \((x,y,z)\), we can represent \(M\) for Fig. \ref{fig:ex2d} as follows:

\[
\begin{bNiceMatrix}[first-row, first-col]
\CodeBefore
\rectanglecolor{blue!15}{3-1}{3-4}
\rectanglecolor{blue!15}{1-3}{4-3}
\Body  
    & p_1 & p_2 & p_3 & p_4 \\
    p_1 & \sum_{2,3,4} P(1,j) & -P(1,2) & -P(1,3) & -P(1,4) \\
    p_2 & -P(2,1) & \sum_{1,4} P(2,j) & 0 & -P(2,4) \\
    p_3 & -P(3,1) & 0 & \sum_{1,4} P(3,j) & -P(3,4) \\
    p_4 & -P(4,1) & -P(4,2) & -P(4,3) & \sum_{1,2,3} P(4,j) \\
\end{bNiceMatrix}
\]

where \(P(i,j) = (p_i - p_j)(p_i - p_j)^T\).

If the rank of \(M\) satisfies the Maxwell counting criteria (3D) for the rank of the symmetric matrix, which is 
\[
\rank M = dn - \binom{d+1}{2} \quad \text{for } d=3,
\]
then the graph is infinitesimally rigid. Moreover, the zeros within the matrix indicate unconnected nodes, allowing us to count the number of symmetric zeros to assess how far the graph is from being a complete graph, where every vertex is connected to every other vertex.

If the graph has the same number of edges as a complete graph, it implies a unique realization (up to reflection). In this case, we can estimate the number of realizations to be \(2\) because there is one missing node (two zeros) from the complete graph. In Fig. \ref{fig:ex2d} (a), if we consider \(p_3\) with the edges \((p_1, p_3)\) and \((p_3, p_4)\) as \(I_2\) and ignore the diagonal part of the matrix, the blue shaded area in \(M\) represents the \(I_2\) part. Therefore, connected triangles can be decomposed into a single triangle plus connected \(I_2\)s. Each \(I_2\) can be positioned ``up" or ``down," meaning that each \(I_2\) represents \(2\) realizations if we fix the orientation of the triangle. Thus, for Fig. \ref{fig:ex2d} (b), we get \(2\) realizations up to reflection.

In this case, the vertices in the triangle \((p_1, p_2, p_3)\) can be regarded as ``pinned," which means assigning fixed coordinate vectors with no changes in their positions. The remainder of the graph is termed a pinned graph, focusing only on \(p_4\) and the edges connected to pinned vertices. Furthermore, the graph is described as pinned rigid if it is rigid, considering the connections of pinned vertices as forming a complete graph.
\subsubsection{Systematic Construction Methods}
Using Cayley-Menger determinants, we can verify whether a given construction yields unique realization through distance matrix completion. For a graph with $n$ vertices in $\mathbb{R}^d$, where $d$ vertices are pinned (fixed), the number of possible realizations with proper connection and algebraically independent edge lengths is typically bounded by:
\begin{equation}
    2^{n-d} \leq N \leq 4^{n-d}
\end{equation}
Here we present methods for constructing pinned rigid graphs with controlled numbers of realizations.

\paragraph{General Construction Strategies}
For graphs in $\mathbb{R}^d$, we have three main approaches: Merge rigid graphs using a rigid hyperplane (face in 3D); combine rigid graphs with pinned rigid graphs (fixed vertices); and add $I_n$ to a rigid graph where $n \geq d$.

\paragraph{Unique Realization}
For unique realization, we can construct surfaces by combining tetrahedra and fixing boundary points such that: $|E| \geq \binom{m+1}{2}$ where $m$ is the number of unpinned vertices; and $E$ excludes edges between pinned vertices (computable from known coordinates).

This condition becomes both necessary and sufficient for uniqueness when edge lengths are algebraically independent and no special geometric configurations occur, such as collinear or coplanar vertices. Without algebraic independence, structures with fewer edges might still have unique realizations due to special geometric relationships. For example, symmetric configurations or specific metric conditions might force uniqueness with fewer constraints.

\paragraph{Topological Structures}
These construction strategies yield different types of structures: Discrete Strip (genus 0): Sequential construction from one or both ends (Fig. \ref{fig:strip-annulus} (a) or Fig. \ref{fig:marching}), and Discrete Annulus (genus 1): Direction-independent construction maintaining rigidity (Fig. \ref{fig:strip-annulus} (b) or Fig. \ref{fig:grid}).

\paragraph{Local and Global Solvability}
The existence of a global solution requires local solvability at each construction step. For any substructure constructed based on tetrahedrons, we can use local Cayley-Menger conditions to verify realizability:
Each tetrahedron must satisfy $CM(P_0,...,P_3) = 0$, and connected tetrahedra must satisfy additional conditions $CM(P_0,...,P_4) = 0$.
Only when all local conditions are satisfied can we proceed to analyze global structure. This hierarchical relationship between local and global solvability explains several key phenomena: local geometric incompatibility can prevent global solutions, while the existence of local solutions, though necessary, is not sufficient for global realizability. Furthermore, special configurations may satisfy local conditions while failing to meet global constraints.

However, some minimally rigid structures, such as an octahedron with three pinned points, present an important exception to this sequential analysis. In these cases, although the structure is globally rigid, it cannot be decomposed into sequential tetrahedra, making local Cayley-Menger analysis inapplicable in a sequential manner. Such situations necessitate alternative methods, such as polynomial system solving.

Due to the computational complexity of polynomial solving for large graphs, where each unknown point introduces three variables and a sphere equation (for example, 100 unknown points result in a system of 100 non-linear equations in 300 variables that cannot be solved sequentially), we focus on structures decomposable into tetrahedra (equivalently $d$-simplex for $d$ dimension). This approach allows local computation at each unknown point through tetrahedral constraints.
This restriction can be relaxed through several complementary strategies: adding more pinned points to reduce degrees of freedom, utilizing local tetrahedral decompositions where possible to maintain sequential solvability, and developing hybrid methods that combine local tetrahedral analysis with global constraints.

\begin{figure}[H]
    \centering
        \begin{tikzpicture}[scale=0.50, transform shape]
	\begin{pgfonlayer}{nodelayer}
		\node [style=bluedot] (0) at (-8.5, 3.25) {};
		\node [style=bluedot] (1) at (-8.5, 1.5) {};
		\node [style=blackdot] (2) at (-4, 1.5) {};
		\node [style=bluedot] (3) at (-4, 3.25) {};
		\node [style=blackdot] (4) at (-6.25, 1.5) {};
		\node [style=bluedot] (5) at (-6.25, 3.25) {};
		\node [style=bluedot] (6) at (0.5, 1.5) {};
		\node [style=bluedot] (7) at (0.5, 3.25) {};
		\node [style=blackdot] (8) at (-1.75, 1.5) {};
		\node [style=bluedot] (9) at (-1.75, 3.25) {};
		\node [style=bluedot] (10) at (-8.5, -1.5) {};
		\node [style=bluedot] (11) at (-8.5, -3.25) {};
		\node [style=blackdot] (12) at (-4, -3.25) {};
		\node [style=bluedot] (13) at (-4, -1.5) {};
		\node [style=blackdot] (14) at (-6.25, -3.25) {};
		\node [style=bluedot] (15) at (-6.25, -1.5) {};
		\node [style=bluedot] (16) at (0.5, -3.25) {};
		\node [style=bluedot] (17) at (0.5, -1.5) {};
		\node [style=blackdot] (18) at (-1.75, -3.25) {};
		\node [style=bluedot] (19) at (-1.75, -1.5) {};
		\node [style=none] (20) at (-6.25, 0.75) {\Large $I_3$};
		\node [style=none] (21) at (-4, 0.75) {\Large $I_2$};
		\node [style=none] (22) at (-1.75, 0.75) {\Large $I_4$};
		\node [style=none] (23) at (-6.25, -4.25) {\Large $I_4$};
		\node [style=none] (24) at (-4, -4.25) {\Large $I_3$};
		\node [style=none] (25) at (-1.75, -4.25) {\Large $I_2$};
		\node [style=none] (26) at (-4.75, -2.75) {};
		\node [style=none] (27) at (-3.25, -2.75) {};
		\node [style=none] (28) at (-1, -2.75) {};
		\node [style=none] (29) at (-1, -3.75) {};
		\node [style=none] (30) at (-3.25, -3.75) {};
		\node [style=none] (31) at (-4.75, -3.75) {};
		\node [style=none] (32) at (-4, -0.25) {\huge (a)};
		\node [style=none] (33) at (-4, -5.25) {\huge (b)};
	\end{pgfonlayer}
	\begin{pgfonlayer}{edgelayer}
		\draw [style=greenedge] (0) to (1);
		\draw [style=blackedge](1) to (2);
		\draw [style=blackedge](3) to (2);
		\draw [style=blackedge](0) to (4);
		\draw [style=greenedge] (0) to (5);
		\draw [style=greenedge] (5) to (3);
		\draw [style=blackedge](5) to (4);
		\draw [style=greenedge] (7) to (6);
		\draw [style=greenedge] (9) to (7);
		\draw [style=blackedge](9) to (8);
		\draw [style=blackedge](7) to (8);
		\draw [style=blackedge](8) to (2);
		\draw [style=greenedge] (3) to (9);
		\draw [style=blackedge](8) to (6);
		\draw [style=greenedge] (10) to (11);
		\draw [style=blackedge](11) to (12);
		\draw [style=blackedge](13) to (12);
		\draw [style=blackedge](10) to (14);
		\draw [style=greenedge] (10) to (15);
		\draw [style=greenedge] (15) to (13);
		\draw [style=blackedge](15) to (14);
		\draw [style=greenedge] (17) to (16);
		\draw [style=greenedge] (19) to (17);
		\draw [style=blackedge](19) to (18);
		\draw [style=blackedge](18) to (12);
		\draw [style=greenedge] (13) to (19);
		\draw [style=blackedge](18) to (16);
		\draw [style=blackedge](14) to (13);
		\draw [gray, thick, pattern=north west lines, pattern color=gray] (26.center) to (27.center)--(27.center) to (28.center)--(26.center) to (31.center)--(31.center) to (30.center)--(30.center) to (29.center)--(29.center) to (28.center);
	\end{pgfonlayer}
\end{tikzpicture}
\begin{tikzpicture}[scale=0.50, transform shape]
	\begin{pgfonlayer}{nodelayer}
		\node [style=blackdot] (12) at (-4, -3.25) {};
		\node [style=bluedot] (13) at (-4, -1.5) {};
		\node [style=bluedot] (14) at (-6.25, -3.25) {};
		\node [style=bluedot] (15) at (-6.25, -1.5) {};
		\node [style=bluedot] (16) at (0.5, -3.25) {};
		\node [style=bluedot] (17) at (0.5, -1.5) {};
		\node [style=blackdot] (18) at (-1.75, -3.25) {};
		\node [style=bluedot] (19) at (-1.75, -1.5) {};
		\node [style=none] (24) at (-4, -4.25) {\Large $I_3$};
		\node [style=none] (25) at (-1.75, -4.25) {\Large $I_2$};
		\node [style=none] (33) at (-4, -5.25) {\huge (c)};
		\node [style=none] (34) at (-6.25, -4.25) {\Large $I_4$};
	\end{pgfonlayer}
	\begin{pgfonlayer}{edgelayer}
		\draw [style=blackedge] (13) to (12);
		\draw [style=greenedge] (15) to (13);
		\draw [style=greenedge] (15) to (14);
		\draw [style=greenedge] (17) to (16);
		\draw [style=greenedge] (19) to (17);
		\draw [style=blackedge] (19) to (18);
		\draw [style=blackedge] (18) to (12);
		\draw [style=greenedge] (13) to (19);
		\draw [style=blackedge] (18) to (16);
		\draw [style=blackedge] (14) to (12);
	\end{pgfonlayer}
\end{tikzpicture}
    \caption{(a) has a solution that can be thought of as a connection of three $I_3$ graphs. However, (b) will have a degree of freedom similar to the three-bar linkage problem (highlighted with gray pattern) since $I_2$ is not connected to $I_4$. (c) Interpretation of (b). If we assume $I_4$ as a fixed point (because the solution is unique up to reflection), we have two unpinned points ($2\cdot 3=6$ in $\mathbb{R}^3$) with five Euclidean distance equations.}
    \label{fig:two-strip}
\end{figure}
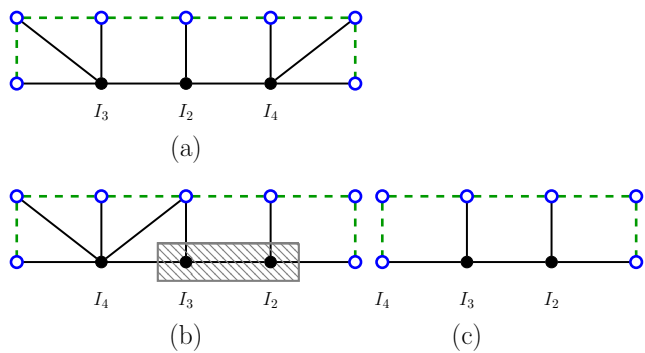
\begin{example}[Construction and Solvability]
We examine two cases in $\mathbb{R}^3$ that illustrate the relationship between local and global solvability:\\
\textbf{Case 1: Finite Realizations.}\\
When $I_2$ connects to $I_3$ and $I_4$, it forms the equivalent of three $I_3$ structures and yields $\leq 2^3$ embeddings when properly constrained.\\
\textbf{Case 2: Continuous Deformation.}\\
When $I_3$ connects to $I_4$ and adjacent $I_2$, we observe a situation involving five distances for two points (six variables), which results in continuous deformation.    
\end{example}
For instance, in three-dimensional space as shown in Fig.~\ref{fig:two-strip}, if $I_2$ is connected to one $I_3$ and one $I_4$, it can be conceptualized as three $I_3$ graphs. In Fig.~\ref{fig:two-strip} (a), a configuration is illustrated with ($\leq 2^3$) possible embeddings. In this scenario, once the positions of the left and right points are determined, the position of the center point can be computed as long as it is connected to at least one pinned point.
Conversely, Fig.~\ref{fig:two-strip} (b) does not have a finite number of solutions and can be interpreted similarly to Fig.~\ref{fig:two-strip} (c). In Fig.~\ref{fig:two-strip} (c), there are two unknown points (represented as black dots) and five known Euclidean distances (sphere equations). Algebraically, this corresponds to five equations with six variables (since there are ($2 \cdot 3$) variables). Consequently, this results in one degree of freedom, indicating that the graph can move freely, assuming all edges are distinct and all vertices are algebraically independent.

Next, we will illustrate an example of gluing two rigid graphs to construct a new rigid graph. 
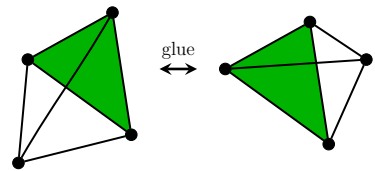
\begin{figure}[H]
    \centering
\begin{tikzpicture}[scale=0.50, transform shape]
	\begin{pgfonlayer}{nodelayer}
		\node [style=blackdot] (0) at (-3, -0.75) {};
		\node [style=blackdot] (2) at (-2.75, 2) {};
		\node [style=blackdot] (3) at (0, 0) {};
		\node [style=blackdot] (4) at (-0.5, 3.25) {};
		\node [style=blackdot] (5) at (2.5, 1.75) {};
		\node [style=blackdot] (6) at (5.25, -0.25) {};
		\node [style=blackdot] (7) at (4.75, 3) {};
		\node [style=blackdot] (8) at (6.25, 2) {};
		\node [style=none] (9) at (0.75, 1.75) {};
		\node [style=none] (10) at (1.75, 1.75) {};
		\node [style=none] (11) at (1.25, 2.25) {};
		\node [style=none] (12) at (1.25, 2.25) {\Large glue};
	\end{pgfonlayer}
	\begin{pgfonlayer}{edgelayer}
		\fill[black!30!green] (2.center) to (3.center) to (4.center) to cycle;
		\fill[black!30!green] (5.center) to (6.center) to (7.center) to cycle;
		\draw [style=blackedge](2) to (0);
		\draw [style=blackedge](2) to (4);
		\draw [style=blackedge](4) to (3);
		\draw [style=blackedge](3) to (0);
		\draw [style=blackedge](3) to (2);
		\draw [style=blackedge,in=60, out=-120] (4) to (0);
		\draw [style=blackedge](5) to (7);
		\draw [style=blackedge](7) to (6);
		\draw [style=blackedge](6) to (5);
		\draw [style=blackedge](5) to (8);
		\draw [style=blackedge](7) to (8);
		\draw [style=blackedge](8) to (6);
		\draw [style=arrow, in=180, out=0, looseness=2.00] (9.center) to (10.center);
	\end{pgfonlayer}
\end{tikzpicture}
\caption{Gluing two tetrahedrons via a green-colored face.}
\label{fig:ge}
\end{figure}
When we connect two tetrahedrons through a green-colored face as shown in Fig. \ref{fig:ge}, this face can function as a pinned vertex. If the three vertices of the shared face are pinned, the remaining structure can be conceptualized as $I_3$. Consequently, the combination of the two tetrahedrons can be viewed as one tetrahedron plus one $I_3$.

\begin{figure}[H]
\begin{center}
	\begin{tabular}{cc}
		\addheight{		\includegraphics[width=3cm]{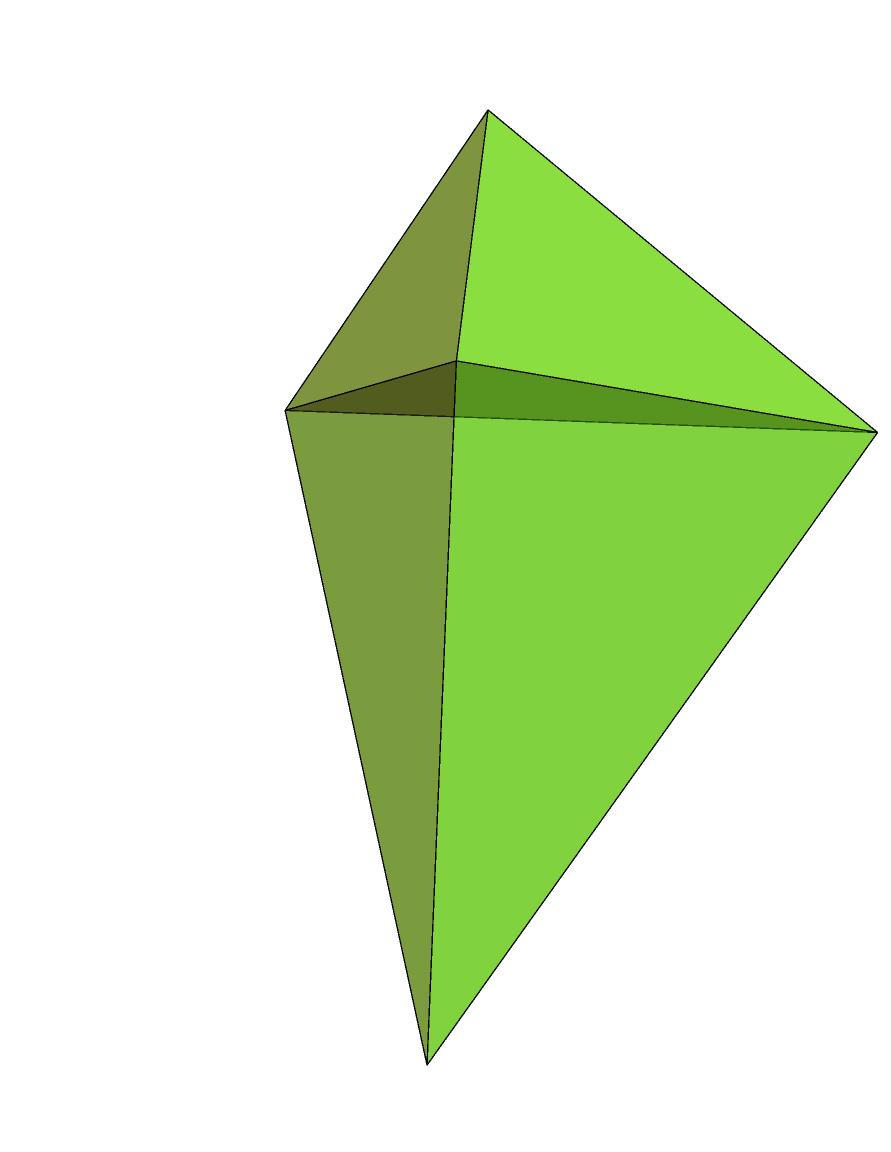}
		} &
		\addheight{	\includegraphics[width=3cm]{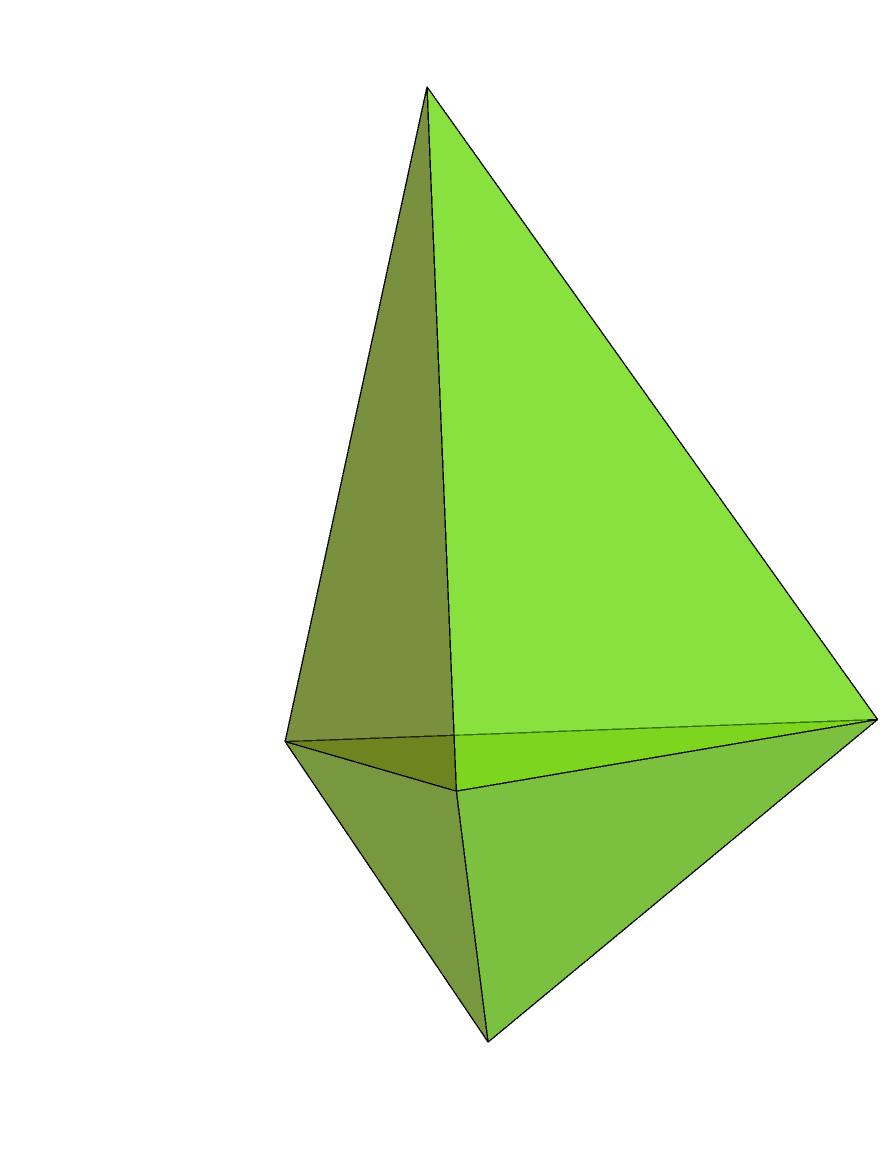}
		} \\
		\small (a) Embedding 1 (unfolded). & (b) Embedding 2 (unfolded).\\
		\addheight{	\includegraphics[width=3cm]{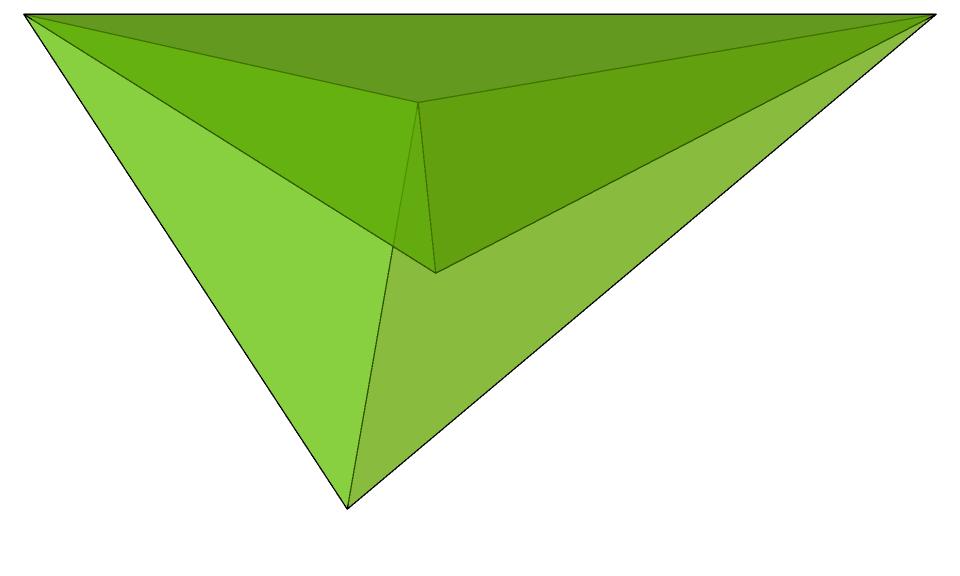}
		}&
		\addheight{	\includegraphics[width=3cm]{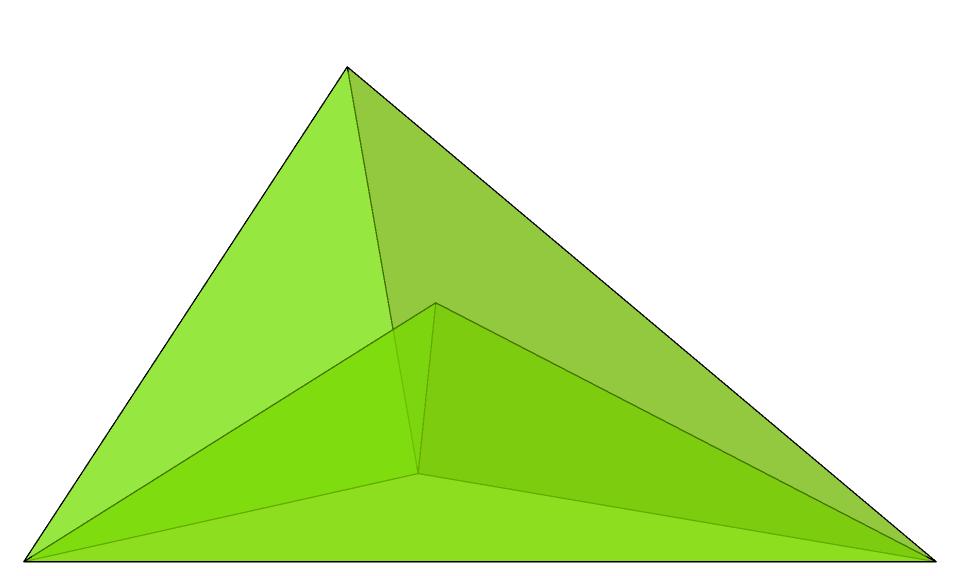}
		} \\
		\small (c) Embedding 3 (folded). & (d) Embedding 4  (folded).\\
	\end{tabular}
	\caption{Four resulting embeddings from Euclidean distance equations of Fig \ref{fig:ge}.}\label{fig:ge2}
\end{center}
\end{figure}

This allows us to estimate the number of embeddings as $2$, accounting for reflections, resulting in a total of $4$ embeddings when reflections are counted separately, as shown in Fig. \ref{fig:ge2}.
 \begin{figure*}
\begin{center}    
    \includegraphics[width=0.85 \textwidth]{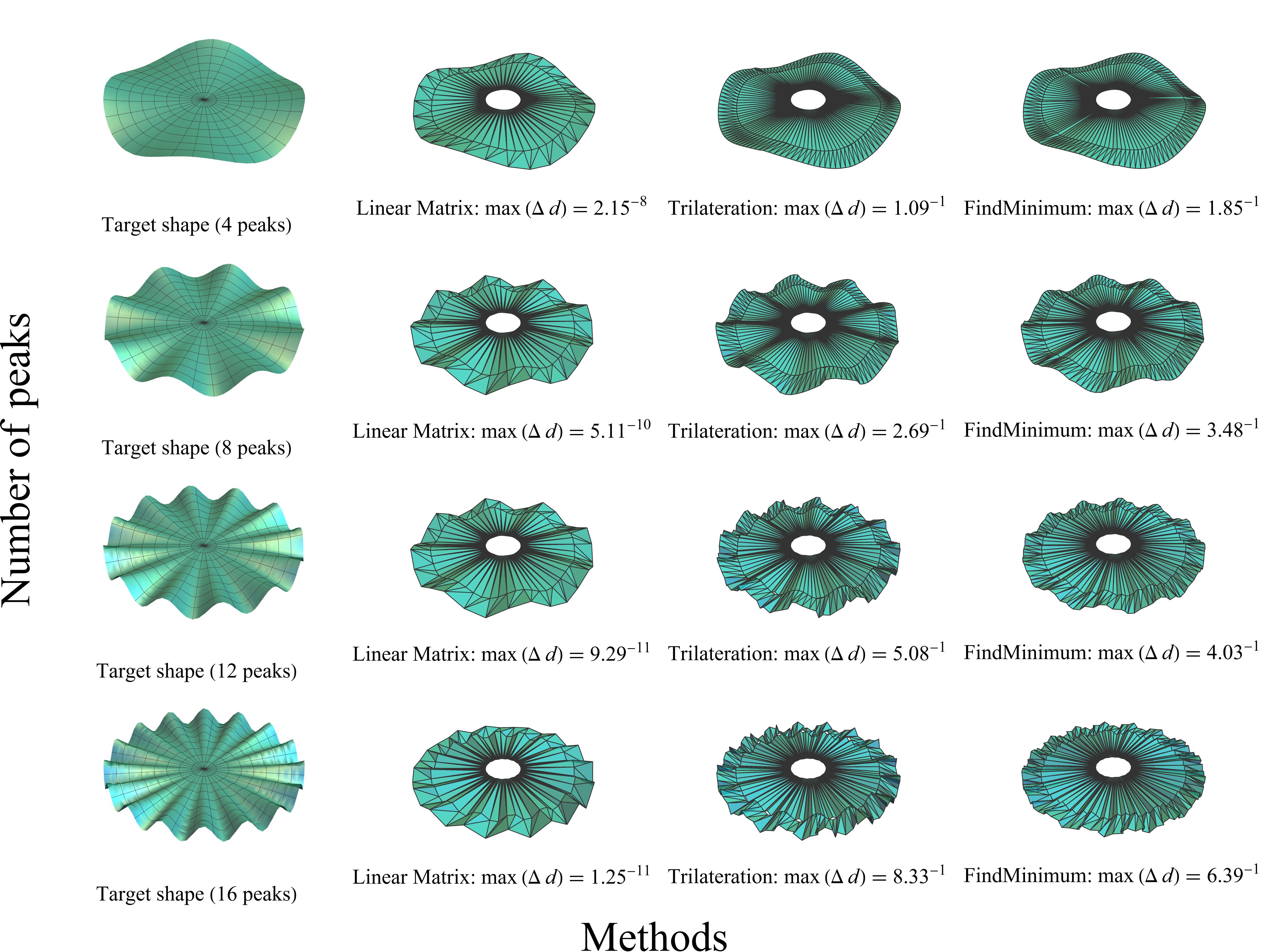}
   
			\caption{{\it Sum of distance energy function $E_{d1}$ by different methods}}
\label{fig:err-method}
\end{center}
\end{figure*}

\subsection{Simulation}\label{sec:simulation}
While our theoretical framework provides bounds on the number of realizations, practical shape reconstruction requires efficient computational methods. Here we present two different grid structures in 3D with their associated reconstruction methods: one yielding $\leq 2^{n-3}$ realizations and another ensuring unique realization.

The first structure, a modified triangulated surface (Fig.~\ref{fig:grid} (b) or \ref{fig:marching} (b)), is based on sequential addition of $I_3$ structures. This configuration yields up to $2^{n-3}$ realizations and requires trilateration for reconstruction. The second structure uses sequential pyramids (Fig.~\ref{fig:grid} (c) or \ref{fig:marching} (c)), based on adding $I_4$ structures, which ensures unique realization and is amenable to linear matrix methods. Both structures avoid crossing edges, allowing construction from flat 2D grids through appropriate edge length constraints.

\subsubsection{Reconstruction Methods}
\begin{figure}[]
    \centering
    \includegraphics[width=0.45 \textwidth]{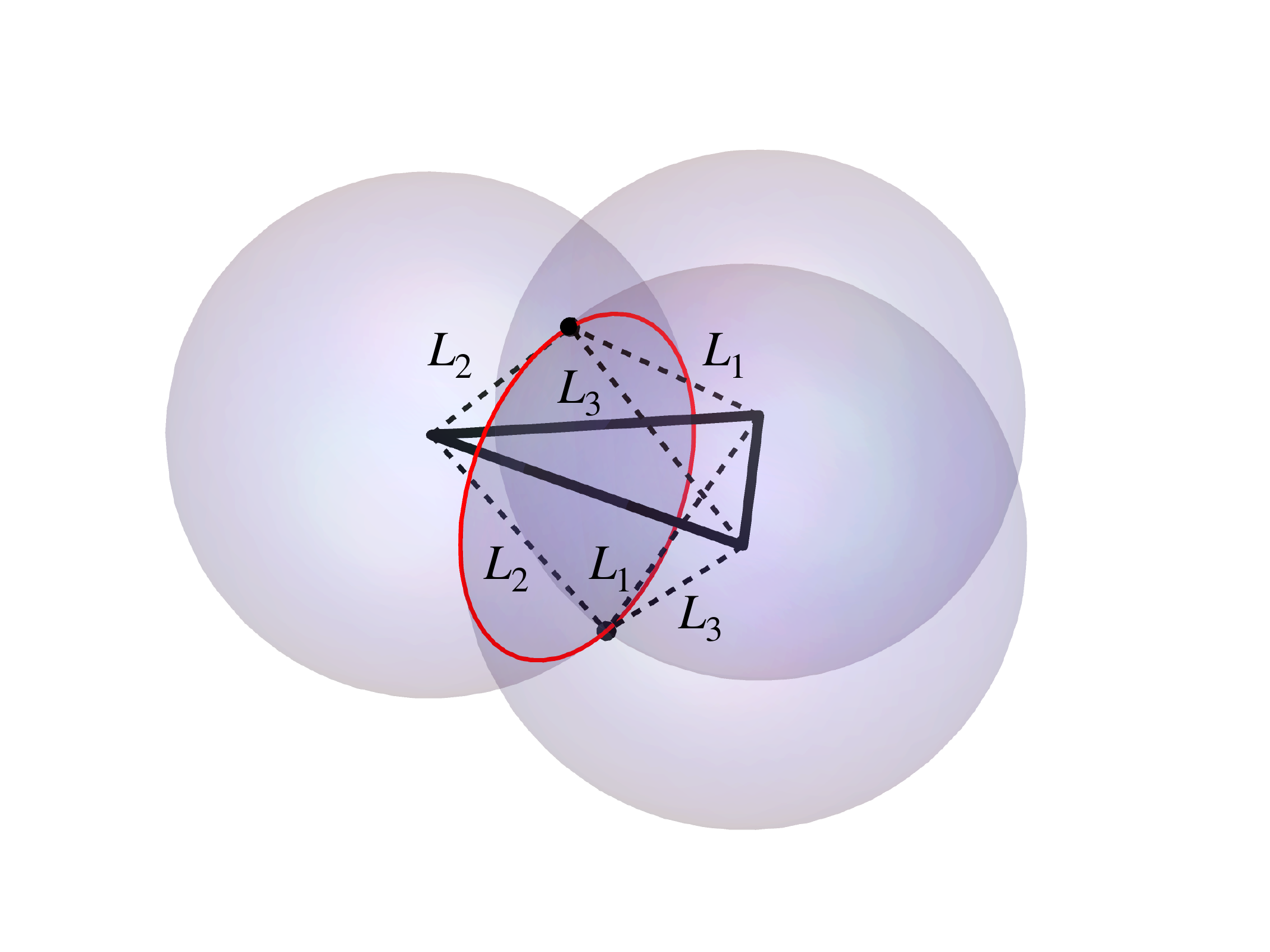}
    \caption{An illustration of trilateration. With fixed edge lengths ($L_1,L_2$, and $L_3$), a vertex joined to three existing vertices can be in one of two places at most.}
    \label{fig:trilaterationsketch}
\end{figure}

Our computations and simulations implement three approaches using R \cite{Rweb} and Mathematica \cite{Mathematica}. The trilateration method, primarily used for the modified triangulated surface, determines vertex positions using three known points and distances. This approach yields two possible positions per vertex (Fig.~\ref{fig:trilaterationsketch}), resulting in $2^m$ total solutions for $m$ reconstructed points. Solutions are selected based on triangle angles relative to previous points.

The linear matrix method (GB in \cite{Liberti2014}), applied to the pyramid structure, uses linearization of distance equations (Eq.~\ref{eq:two-circles2}). While requiring twice the number of fixed points, this method provides unique solutions sequentially and shows better stability. The first point is computed using trilateration with a known solution choice.

In addition to these direct methods, we also explored optimization-based approaches. For the optimization method, we utilized the \texttt{FindMinimum} function in Mathematica \cite{Mathematica} for this section, and the \texttt{optimx} library in R \cite{Rweb} for the appendix. We selected the conjugate gradient option in both cases.
For comparison, we implemented minimization using different energy functions in Mathematica and R:
\[
E_{\text{Mathematica}} = \sum_{i,j} (d(q_i,q_j) - d_{ij})^2
\]
in Mathematica and 
\[
E_{\text{R}} = \sum_{i,j} (d(q_i,q_j)^2 - d_{ij}^2)^2
\]
where $d(x,y)$ is the Euclidean distance function and $d_{ij}$ represents the given distances. The \texttt{FindMinimum} function was initialized with a flat surface configuration $(x,y,z=0)$.

\subsubsection{Test Results}
To evaluate these methods, we used a parametric test function generating surfaces with controlled numbers of peaks:
\[
(x,y,z) = (5 v \cos(u), 5 v \sin(u), (0.3 |v|^2 + 0.2 |v|^4) \sin(au))
\]
where $a = 4, 8, 12, 16$ controls the number of peaks (Fig. \ref{fig:err-method}).
\begin{example}\label{cont-SR}
Before we start to talk about discretized counterparts, we can demonstrate that a unique smooth surface exists given the intrinsic metric and boundary conditions. Consider our target surface defined by the parametric equation with $a=16$:

Suppose we have given metric tensor $E,F,G$ with the boundary condition at $v=1$: $(x,y,z) = (5\cos(u), 5\sin(u), 0.5\sin(16u))$ (corresponding to the boundary of white circular region in Fig. \ref{fig:err-method} for the discrete methods),
\begin{align}
E &= 25v^2 + 16^2(0.3|v|^2 + 0.2|v|^4)^2\cos^2(16u) \\
F &= 16(0.3|v|^2 + 0.2|v|^4)(0.6|v| + 0.8|v|^3)\sin(16u)\cos(16u) \\
G &= 25 + (0.6|v| + 0.8|v|^3)^2\sin^2(16u)
\end{align}
where $E = \mathbf{r}_u \cdot \mathbf{r}_u$, $F = \mathbf{r}_u \cdot \mathbf{r}_v$, and $G = \mathbf{r}_v \cdot \mathbf{r}_v$.

We can determine the surface through the following steps:
1. Based on the radial symmetry and boundary conditions, the general form of our solution must be:
\begin{align}
x &= f_1(v)\cos(u) \\
y &= f_1(v)\sin(u) \\
z &= f_2(v)\sin(16u)
\end{align}
where $f_1(1) = 5$ and $f_2(1) = 0.5$.

2. By examining the form of the metric components and the boundary conditions, we can determine that $f_1(v) = 5v$ and $f_2(v) = (0.3|v|^2 + 0.2|v|^4)$.

3. We verify this solution by checking: At $v=1$: $f_2(1) = 0.3 + 0.2 = 0.5$, and the calculated $E$, $F$, $G$ match our given metric tensor.
   
Since the boundary condition provides information related to second fundamental form (how much it bends on the space), we can reconstruct shape by Gauss-Codazzi. This demonstrates that the case of continuous and bounded surface has a unique solution, which contrasts sharply with the discrete case where multiple realizations can satisfy the same edge length constraints. Another case could be using Euclidean distances instead of the metric tensor, which would make the problem more difficult to solve. However, if we significantly increase the number of sampled edges (for example, using 100 well-distributed Euclidean distances across the parameter space), we can gradually approach the uniqueness of the continuous case due to continuity and by estimating the original functions through their form. Nevertheless, without knowing relations rather than just numerical distances, many different surfaces could satisfy these constraints, illustrating why discrete reconstruction is fundamentally more challenging and prone to multiple solutions.
\end{example}

This fundamental difference between continuous and discrete cases explains why the discrete case shows different behavior among the choice of the grid and methods, as shown in Fig. \ref{fig:err-method}. The linear matrix method demonstrated superior performance by utilizing more fixed points (128 pinned and 96 unpinned points versus 126 pinned and 252 unpinned points for other methods), ensuring better-distributed constraints throughout the structure. When comparing trilateration and optimization approaches, \texttt{FindMinimum} showed smaller errors than trilateration but remained less accurate than the linear matrix method. Interestingly, the linear matrix method's performance improved with increasing peak numbers, as the greater angles between edges enhanced numerical stability.

To test robustness, we added noise to both pinned vertices and true values using normal distribution $(\mu,\sigma)$. The maximum distance error was measured as:
\[
\max \frac{d(v_{i,m}-v_{i,t})}{\sqrt{\Delta}}
\]
where $v_{i,m}$, $v_{i,t}$ are measured and true vertices, and $\Delta$ represents the computation triangles.

Analysis of error distributions (Table \ref{table:1}) revealed that the linear matrix method reduced error from 0.303 to 0.275 for $(\mu,\sigma) = (0, 0.05)$. Despite the different grid sizes, both configurations (96 and 252 unpinned points) showed similar error distributions (Fig.~\ref{fig:err-dist}), with the linear matrix method maintaining error levels comparable to input noise. The method's over-constrained nature prevented error accumulation, demonstrating its robustness to noise.

 \begin{figure}
\begin{center}    
    \includegraphics[width=0.4 \textwidth]{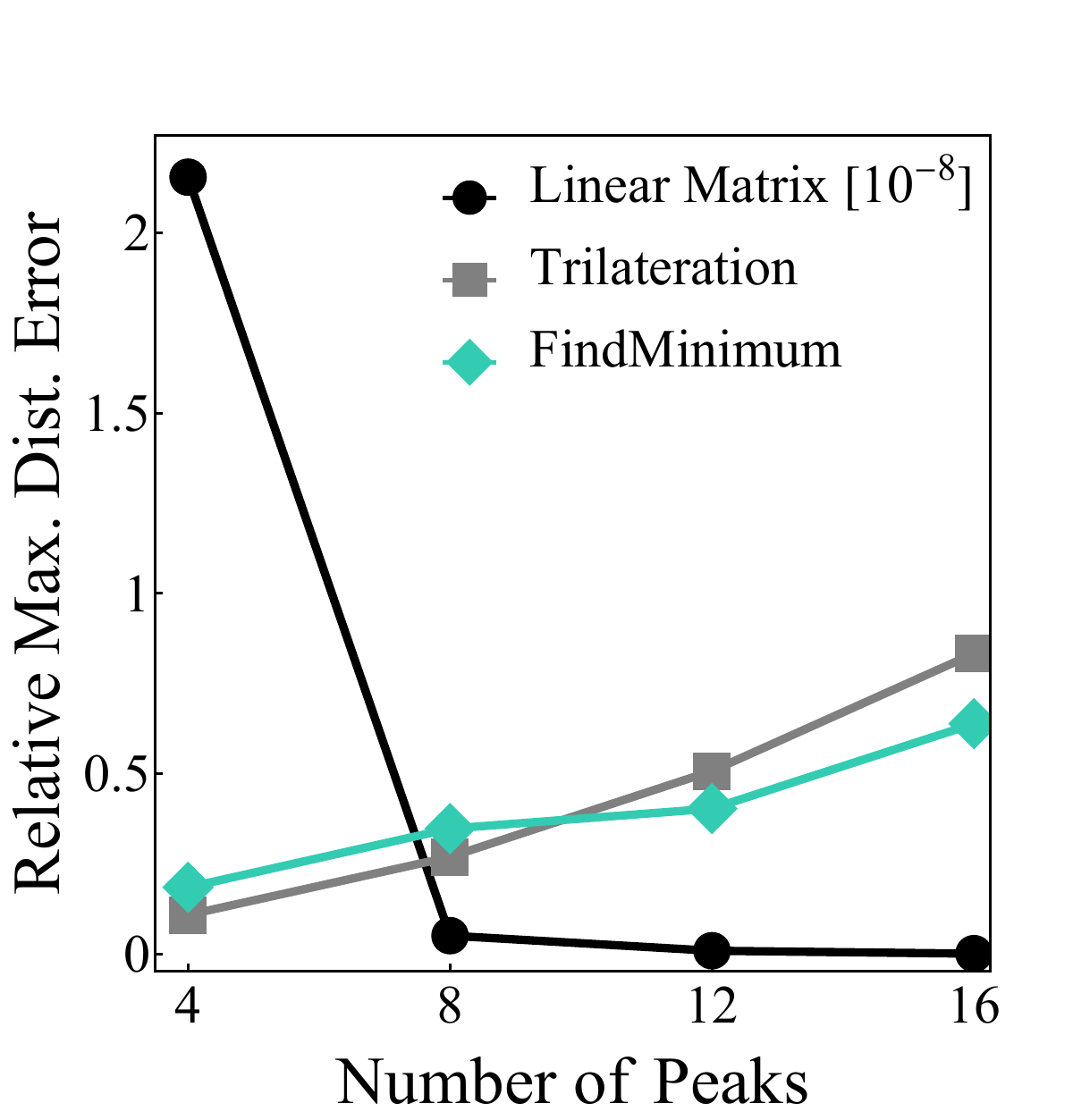}
   
			\caption{{\it Error by Euclidean distance}}
\label{fig:err-method2}
\end{center}
\end{figure}
	\begin{table*}[ht]\caption{Max. Euclidean distance error from noise $(\mu,\sigma)$ (12 peaks)}
	\centering
	\begin{tabular}{|l| c| c| c| c| c|}
		\hline                        
		$(\mu, \sigma)=$& $(0,0.05)$ & $(-0,05,0.05)$ & $(0.05,0.05)$ & $(0,0.07)$ & $(0,0.10)$\\ [0.5ex]
		\hline                 
        Max err. from the noise &  0.303& 0.261 & 0.348& 0.347& 0.394\\	[1ex]      
        Linear matrix &0.275 &0.261 & 0.348& 0.347& 0.394\\	[1ex] 
        \hline
        Max err. from the noise &  0.422 &0.617 & 0.592&0.700 & 0.937\\	[1ex]         
        Trilateration & 1.189 & 1.131&1.270 &1.481 &1.430 \\	[1ex]      
        FindMinimum & 0.788 & 0.838 &0.823 &1.057 & 1.082\\	[1ex]      
		\hline
	\end{tabular}\label{table:1}
\end{table*}
\begin{figure}
\begin{center}    
    \includegraphics[width=0.4 \textwidth]{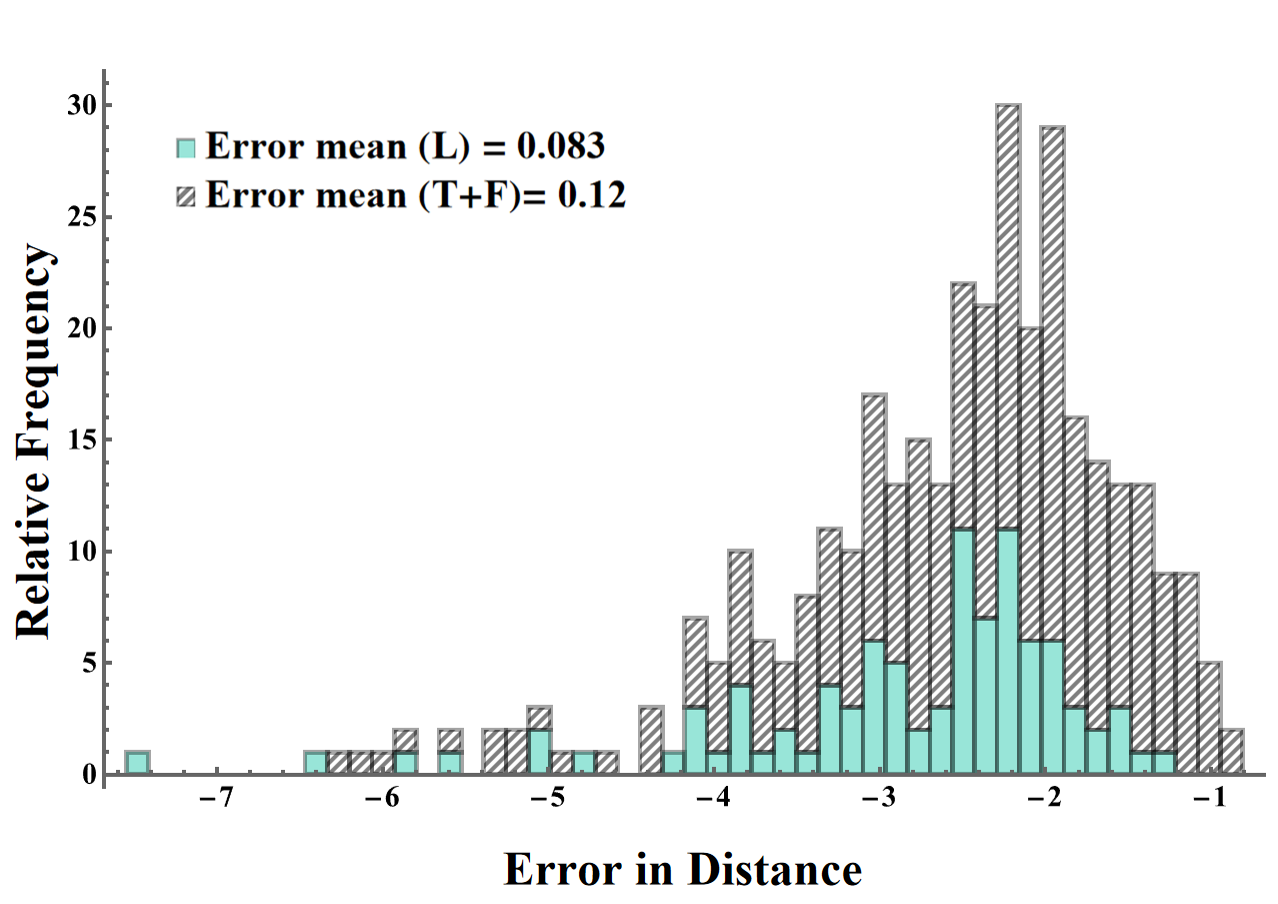}
   
			\caption{{\it Distance error from the noise $(\mu, \sigma)=(0,0.05)$.} Stacked histogram shows distance error by adding random noise. Noisy vertex sets for linear matrix show smaller mean, but overall distributions of two groups are not different.}
\label{fig:err-dist}
\end{center}
\end{figure}

The modified triangulated surface results (shown in Fig. \ref{fig:err-method}, two columns on the right (at 12 peaks and 16 peaks), relative to the target structure) confirm Claim \ref{claim1}. The insufficient constraints on dihedral angles led to increased errors in regions with more peaks, where surface fluctuations exceeded the target surface's peak count. This observation confirms our prediction that incorrect selections among multiple solutions produce unexpected local fluctuations. The addition of targeted constraints, such as slope in $u$ or periodicity conditions (since the discrete boundary values do not help to estimate these conditions solely by coordinate vectors), can help select the desired shape without over-constraining the system. Moreover, these seemingly incorrect solutions can be viewed as physically realizable states arising from an under-constrained system. Similar phenomena appear in continuous systems, where multiple stable configurations in isometric immersions can arise naturally when minimizing elastic energy without considering bending contributions \cite{Gemmer2016}. 

The improved performance of the linear matrix method with increased peaks, as shown in Fig. \ref{fig:err-method2}, provides insights about the importance of a well-constrained system for fluctuating surfaces. These results demonstrate that proper constraint selection and method choice, related to the number of realizations from given constraints, can significantly impact reconstruction accuracy, particularly in regions of high geometric complexity.
\section{Discussion}
Our analysis of discrete surface embeddings demonstrates a fundamental discrepancy between the continuous and discrete representations of surfaces in physical systems. While continuous surfaces with well-behaved metrics typically admit unique isometric embeddings under suitable boundary conditions, their discrete counterparts-networks of vertices connected by edges-can exhibit multiple distinct realizations despite identical edge length constraints. The methods we have developed address this challenge through systematic construction of mesh topologies that yield controllable numbers of embeddings, computational techniques that efficiently determine these embeddings, and geometric insights that relate local fluctuations to constraint insufficiency. These findings bridge theoretical rigidity analysis with practical implementation strategies for applications ranging from 4D printing to mechanical meta-materials.

While our construction method focused primarily on edge lengths and distance matrices, the framework can be extended to broader physical systems. In particular, the geometric constraints represented by edge lengths can be generalized to force-based constraints, provided they can be formulated in terms of linear operations on vertex coordinates. For example, in elastic systems, the stiffness matrix $M$ relates forces and displacements via $M X = F$. At equilibrium ($F = 0$), this relationship determines allowable coordinate vectors, analogous to our distance-based constraints. This principle extends to various physical systems where forces can be expressed in matrix form.

\begin{example}
Consider a system of particles interacting through gravitational forces:
\begin{equation}
F_{ij} = G\frac{m_im_j}{d_{ij}^2}
\end{equation}

For particles with equal masses and fixed distances, we can write the force balance for each particle $i$:
\begin{equation}
\sum_{j\neq i} \frac{G m^2}{d_{ij}^2}(\mathbf{x}_j - \mathbf{x}_i) = 0
\end{equation}

This can be written in matrix form for a system of 4 particles:
\[
\begin{pmatrix}
-\sum_j \frac{1}{d_{1j}^2} & \frac{1}{d_{12}^2} & \frac{1}{d_{13}^2} & \frac{1}{d_{14}^2} \\
\frac{1}{d_{12}^2} & -\sum_j \frac{1}{d_{2j}^2} & \frac{1}{d_{23}^2} & \frac{1}{d_{24}^2} \\
\frac{1}{d_{13}^2} & \frac{1}{d_{23}^2} & -\sum_j \frac{1}{d_{3j}^2} & \frac{1}{d_{34}^2} \\
\frac{1}{d_{14}^2} & \frac{1}{d_{24}^2} & \frac{1}{d_{34}^2} & -\sum_j \frac{1}{d_{4j}^2}
\end{pmatrix}
\begin{pmatrix}
\mathbf{x}_1 \\
\mathbf{x}_2 \\
\mathbf{x}_3 \\
\mathbf{x}_4
\end{pmatrix} = 0
\]
where $Gm^2$ has been absorbed into the distances for simplicity.
\end{example}

This formulation preserves the key features of our framework while allowing for physically meaningful configurations with varying inter-particle distances. The efficiency of computation can be improved when some particles are fixed, enabling sequential solution strategies similar to our previous constructions.

Beyond the specific methods presented, our approach reveals fundamental insights about the relationship between discrete and continuous descriptions of physical systems. The discrepancy between these representations has implications for the broader field of computational mechanics, where discretization schemes often implicitly assume that the discrete system will faithfully represent the continuous one. Our findings suggest that careful consideration of constraint topology is essential for reliable physical simulation, particularly in applications where solution uniqueness matters. This perspective may help inform discretization choices in fields ranging from finite element analysis to physical simulation for computer graphics, where unexpected solution multiplicity could lead to numerical instabilities or physically implausible results.

\subsection{Limitations and challenges}
Our approach to discrete surface embeddings, while effective for controlling solution multiplicity and ensuring reliable reconstruction, faces several important limitations. Computational complexity becomes a significant barrier for very large meshes. The sequential nature of our trilateration method means errors can accumulate for structures with thousands of vertices, while the linear matrix approach requires solving larger systems of equations as mesh size increases. This scalability limitation becomes particularly pronounced for real-time applications or when working with dynamically changing structures.

Special geometric configurations present unique challenges. When points are coplanar or nearly coplanar, numerical instability can lead to significant errors in both trilateration and linear matrix methods. These degenerate cases require special handling through robust numerical techniques or alternative formulations. Similarly, our methods assume algebraic independence of edge lengths, which may not hold in practice due to inherent geometric constraints or symmetries in physical systems. 

Our current framework does not directly address the interplay between discrete embedding multiplicity and physical stability. While we can control the number of potential realizations, determining which solution represents the physically preferred state (e.g., minimum energy configuration other than elasticity) requires additional analysis. This limitation is particularly relevant for applications like 4D printing, where predicting the actual physical outcome among mathematically valid embeddings remains challenging. However, things get easier when we do not strictly require distance constraints only, we can add additional parameters to stabilize and solve problems we discussed.

\subsection{Generalization of methods}
The framework developed in this paper extends naturally to higher-dimensional spaces and more general constraint systems. For embeddings in $\mathbb{R}^d$ with $d > 3$, our analysis of In structures generalizes, with each additional dimension requiring corresponding adjustments to the minimal rigidity conditions. The B\'ezout theorem-based counting methods remain applicable, though the algebraic complexity increases substantially.

Time-dependent systems represent another promising direction for generalization. By treating distances as functions of time rather than fixed constraints, our approach could model dynamic structures such as deployable mechanisms or shape-changing materials. This would require extending the rigidity analysis to account for velocity constraints and time-varying rigidity matrices.

Our construction methods also generalize to non-Euclidean target spaces. For applications involving embeddings into spherical or hyperbolic spaces, the distance constraints would be replaced by corresponding geodesic distances. The sequential computation techniques developed here could be adapted to these settings by substituting appropriate distance functions and modifying the geometric operations accordingly. This generalization opens potential applications in computational geometry, computer graphics, and computational relativity.

\subsection{Extended applications}
Beyond the core applications discussed in our results, the methods developed here have potential impact across diverse domains. In computational biology, our approach could address protein folding problems, where the native structure of a protein must be reconstructed from a limited set of distance constraints measured through nuclear magnetic resonance or cross-linking experiments. The ability to analyze solution multiplicity could help identify proteins with multiple stable conformations.

Our analysis of how constraint topology affects solution space could inform more efficient simulation algorithms that avoid unnecessary computation while maintaining physical plausibility. Our systematic approach to controlling solution multiplicity could help develop more reliable design methodologies for complex structural systems.


\section{Appendixes}
\subsection{Detailed information about computational methods}
In this section, we include detailed information on how to construct the grid and perform computations using trilateration or linear matrix methods.

\subsubsection{Designing grid for specific computational method}
There are many different ways to construct grids for the linear matrix and trilateration methods. However, we focused on non-overlapping edges, with modifications based on a typical triangulated surface, as shown in Fig. \ref{fig:marching}(a). 

For trilateration, we swap one edge to the opposite side to connect it to three pinned points. The next point is computed using the previous point along with two pinned points, so the direction of computation must be fixed. In the linear matrix method, this swap operation occurs for every other edge, so each point is connected to three pinned points. This implies that the direction of computation does not matter. Therefore, by computing the mean value from several different sets in different directions, error inflation can be stabilized.

\begin{figure}[h]
    \centering
    \begin{tabular}{c c}
        \includegraphics[width=0.2 \textwidth]{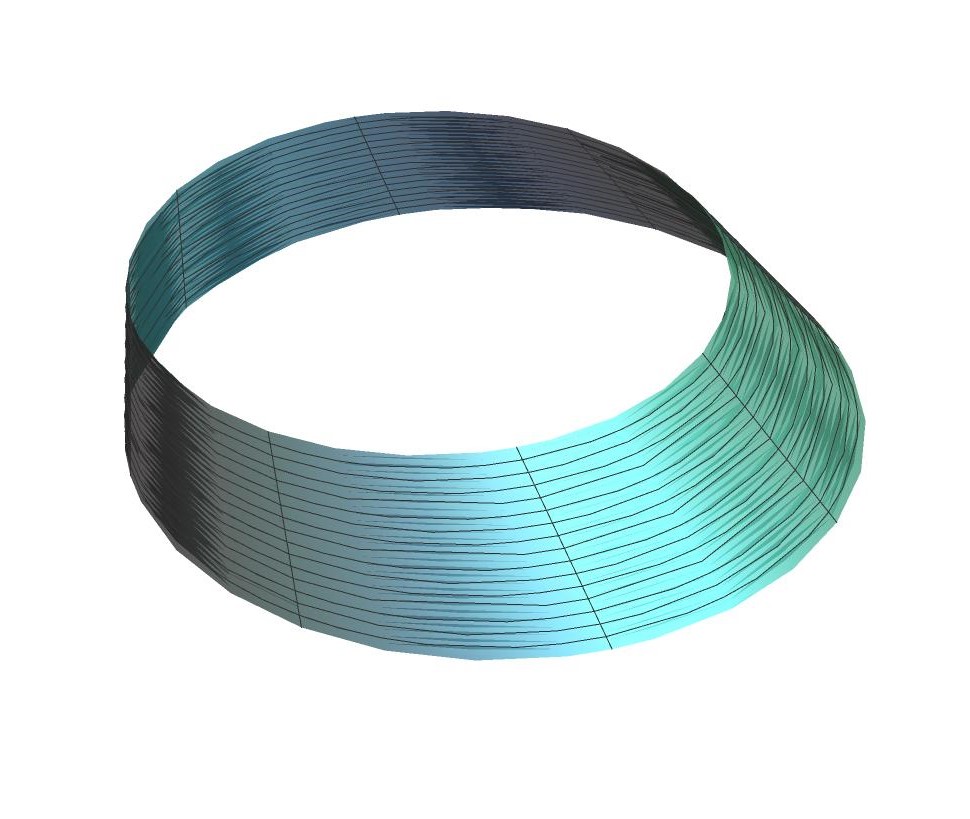} & 
        \includegraphics[width=0.2 \textwidth]{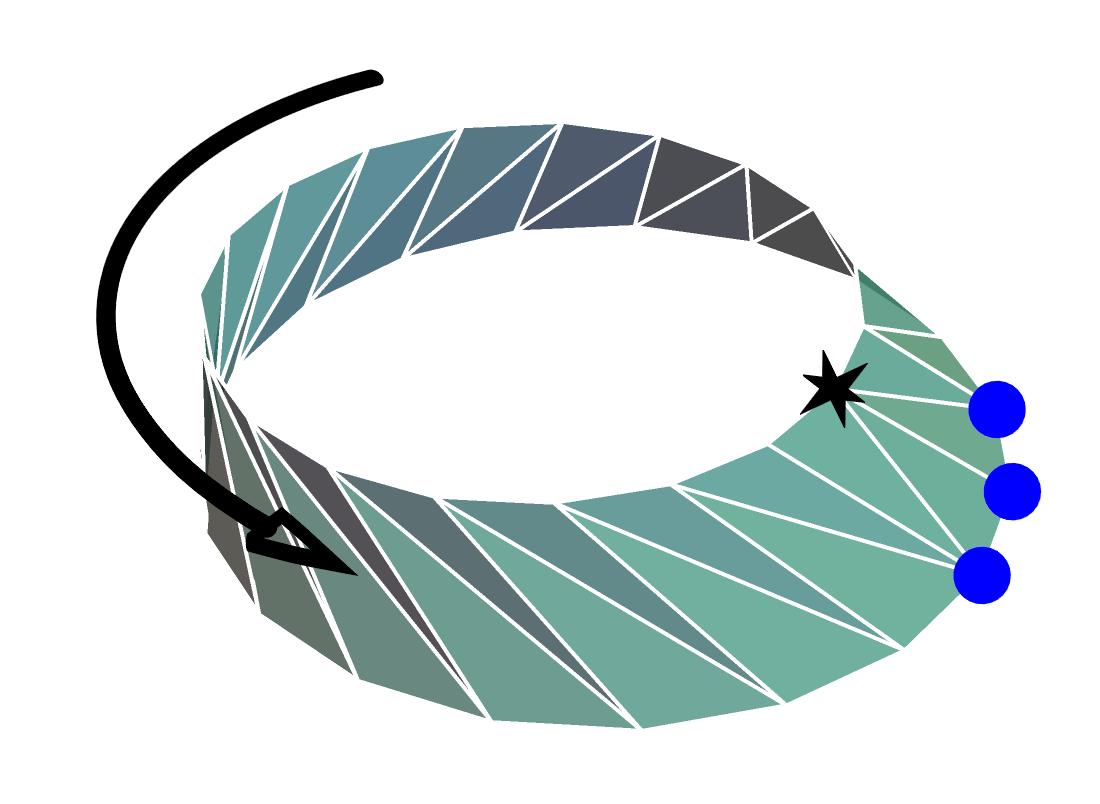} \\
        (a) annulus type surface & (b) grid for trilateration \\
        \includegraphics[width=0.2 \textwidth]{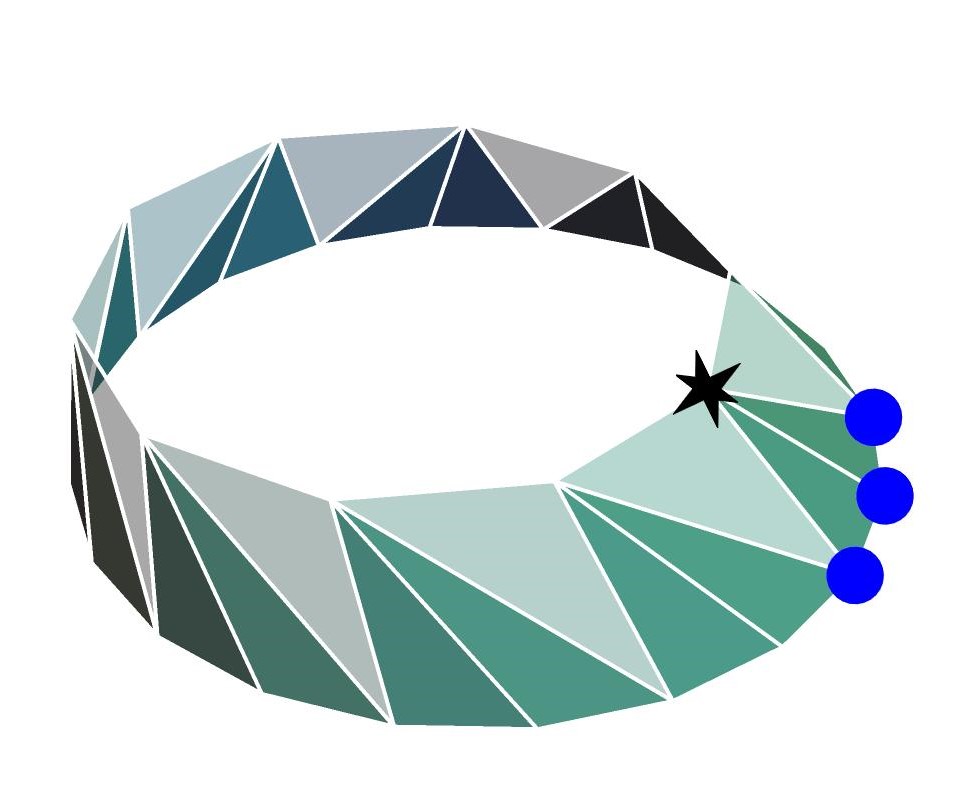} \\
        (c) grid for linear matrix
    \end{tabular}
    \caption{Grid construction for each method: (a) annulus type surface, (b) grid construction example for trilateration, and (c) grid construction example for linear matrix. The star-shaped points represent points to be calculated, and the blue dots represent fixed points used for the calculations.}
    \label{fig:grid}
\end{figure}

\begin{figure}[h]
    \centering
    \begin{tabular}{c}
        \includegraphics[width=0.4 \textwidth]{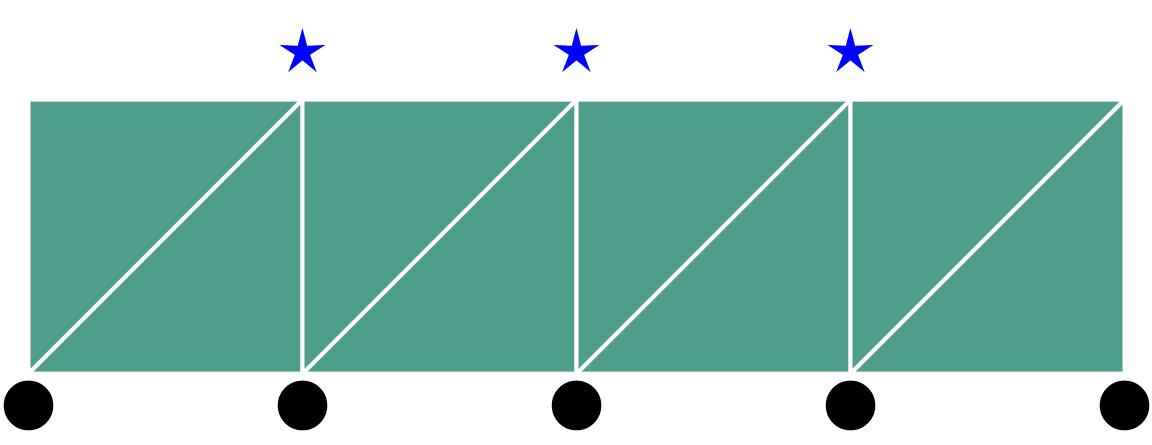} \\
        (a) Regular triangulation \\
        \includegraphics[width=0.4 \textwidth]{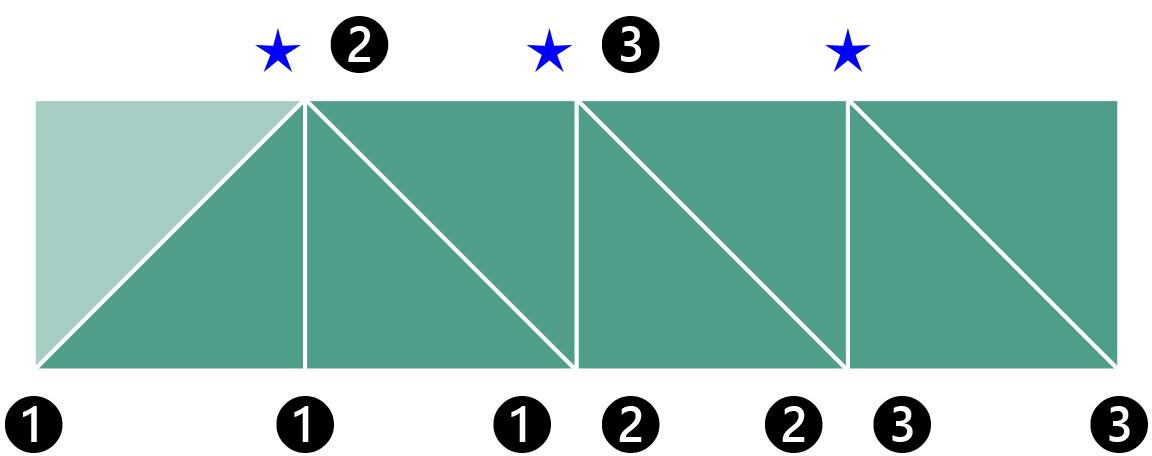} \\
        (b) Grid structure for trilateration \\
        \includegraphics[width=0.4 \textwidth]{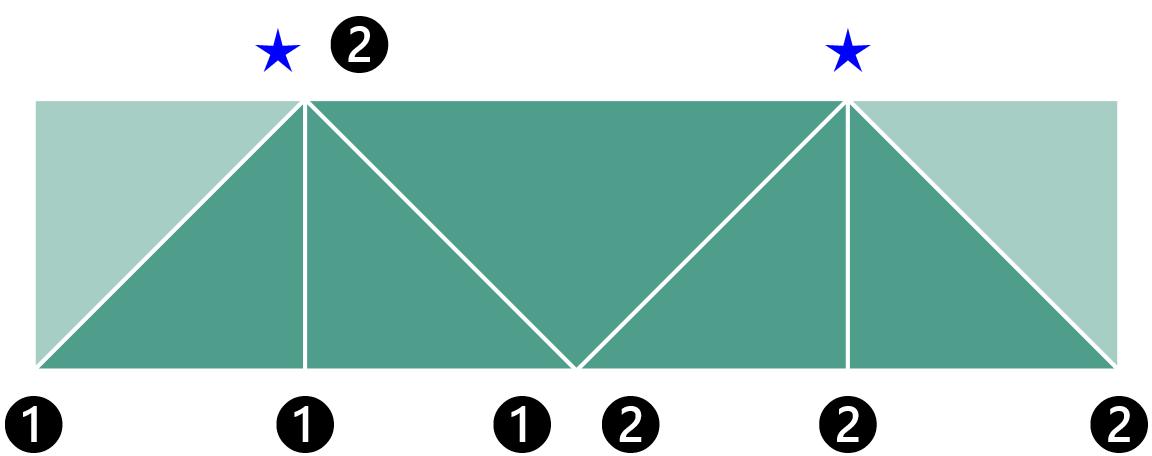} \\
        (c) Grid structure for the linear matrix method
    \end{tabular}
    \caption{Planar view of a 3D strip. (a) Regular triangulation where black dots represent pinned points and star-shaped vertices represent unknown points (vertices to be computed). (b) Grid structure for trilateration. (c) Grid structure for the linear matrix method.}
    \label{fig:marching}
\end{figure}
\subsubsection{Detailed description of marching methods}
Consider 3D strips as shown in Fig. \ref{fig:marching}(a). As mentioned earlier, the grid structure for trilateration can be created by swapping the direction of one edge compared to the others. For example, in Fig. \ref{fig:marching}(b), the left-most edge has a different direction, so the first star-shaped vertex can be connected to three points (marked with circled number 1). The next star-shaped point can then be estimated using the previous point and two pinned points (marked with circled number 2). 

For the linear matrix method, we apply the same strategy, except that each point is connected to three pinned points, assigning two triangles to each star-shaped point. In Figure \ref{fig:marching}(c), each numbered vertex is computed independently using four reference points, allowing for parallel computation and reducing error propagation.

\subsubsection{Computational algorithms}
The following algorithms detail the computational methods used for point determination. Each represents a different approach to solving for unknown vertex positions.

\begin{algorithm}[H]
\caption{Trilateration}
\begin{algorithmic}[1]
\Require 3D position vectors $p_1$, $p_2$, $p_3$ with corresponding Euclidean distances $r_1$, $r_2$, $r_3$ to unknown point $x$
\Ensure Two possible solutions for point $x$
\State $e_x \gets (p_2 - p_1)/|p_2 - p_1|$
\State $i \gets \text{dot}(e_x, p_3 - p_1)$
\State $t \gets (p_3 - p_1 - e_x \cdot i)$
\State $e_y \gets t/|t|$
\State $e_z \gets \text{cross}(e_x, e_y)$
\State $d \gets |p_2 - p_1|$
\State $j \gets \text{dot}(e_y, p_3 - p_1)$
\State $x \gets (r_1^2 - r_2^2 + d^2)/(2 \cdot d)$
\State $y \gets (r_1^2 - r_3^2 - 2 \cdot i \cdot x + i^2 + j^2)/(2 \cdot j)$
\State $z \gets \sqrt{r_1^2 - x^2 - y^2}$
\State $\text{ans1} \gets p_1 + x \cdot e_x + y \cdot e_y + z \cdot e_z$
\State $\text{ans2} \gets p_1 + x \cdot e_x + y \cdot e_y - z \cdot e_z$
\State \Return ans1, ans2
\end{algorithmic}
\end{algorithm}

Note that $|\cdot|$ denotes the norm function, while $\text{dot}(\cdot)$ and $\text{cross}(\cdot)$ represent the inner product and cross product, respectively.

\begin{algorithm}[H]
\caption{Linear Matrix}
\begin{algorithmic}[1]
\Require 3D position vectors $p_1$, $p_2$, $p_3$, $p_4$ with corresponding Euclidean distances $r_1$, $r_2$, $r_3$, $r_4$ to unknown point $x$
\Ensure Solution for unknown point $x$
\State $b_1 \gets r_1^2 - r_4^2 - |p_1|^2 + |p_4|^2$
\State $b_2 \gets r_2^2 - r_4^2 - |p_2|^2 + |p_4|^2$
\State $b_3 \gets r_3^2 - r_4^2 - |p_3|^2 + |p_4|^2$
\State $a_{1x} \gets p_{4x} - p_{1x}$
\State $a_{1y} \gets p_{4y} - p_{1y}$
\State $a_{1z} \gets p_{4z} - p_{1z}$
\State $a_{2x} \gets p_{4x} - p_{2x}$
\State $a_{2y} \gets p_{4y} - p_{2y}$
\State $a_{2z} \gets p_{4z} - p_{2z}$
\State $a_{3x} \gets p_{4x} - p_{3x}$
\State $a_{3y} \gets p_{4y} - p_{3y}$
\State $a_{3z} \gets p_{4z} - p_{3z}$
\State Define $B$ as column vector $B = [b_1/2, b_2/2, b_3/2]^T$
\State Define $A$ as $3 \times 3$ matrix $A = 
\begin{bmatrix} 
a_{1x} & a_{1y} & a_{1z} \\
a_{2x} & a_{2y} & a_{2z} \\
a_{3x} & a_{3y} & a_{3z}
\end{bmatrix}$
\State Solve linear system $Ax = B$ for unknown coordinates $x$
\State \Return $x$
\end{algorithmic}
\end{algorithm}

\subsubsection{Comparison of methods}
While trilateration provides two possible solutions per vertex (requiring additional criteria to select the correct one), the linear matrix method yields a unique solution directly. However, the linear matrix method requires four reference points instead of three, making it more constrained but less prone to error accumulation in extended structures.

In Figure \ref{fig:marching}(b), the numbered vertices indicate the computation sequence for trilateration. Each numbered vertex is computed using three previously determined points, with the sequence ensuring that each new vertex connects to at least one previously computed vertex and two pinned vertices. This sequential approach means errors can potentially accumulate as the computation progresses through the mesh.

\subsubsection{Possible issues with marching methods}
\begin{figure}[H]
    \centering
    \includegraphics[width=0.4 \textwidth]{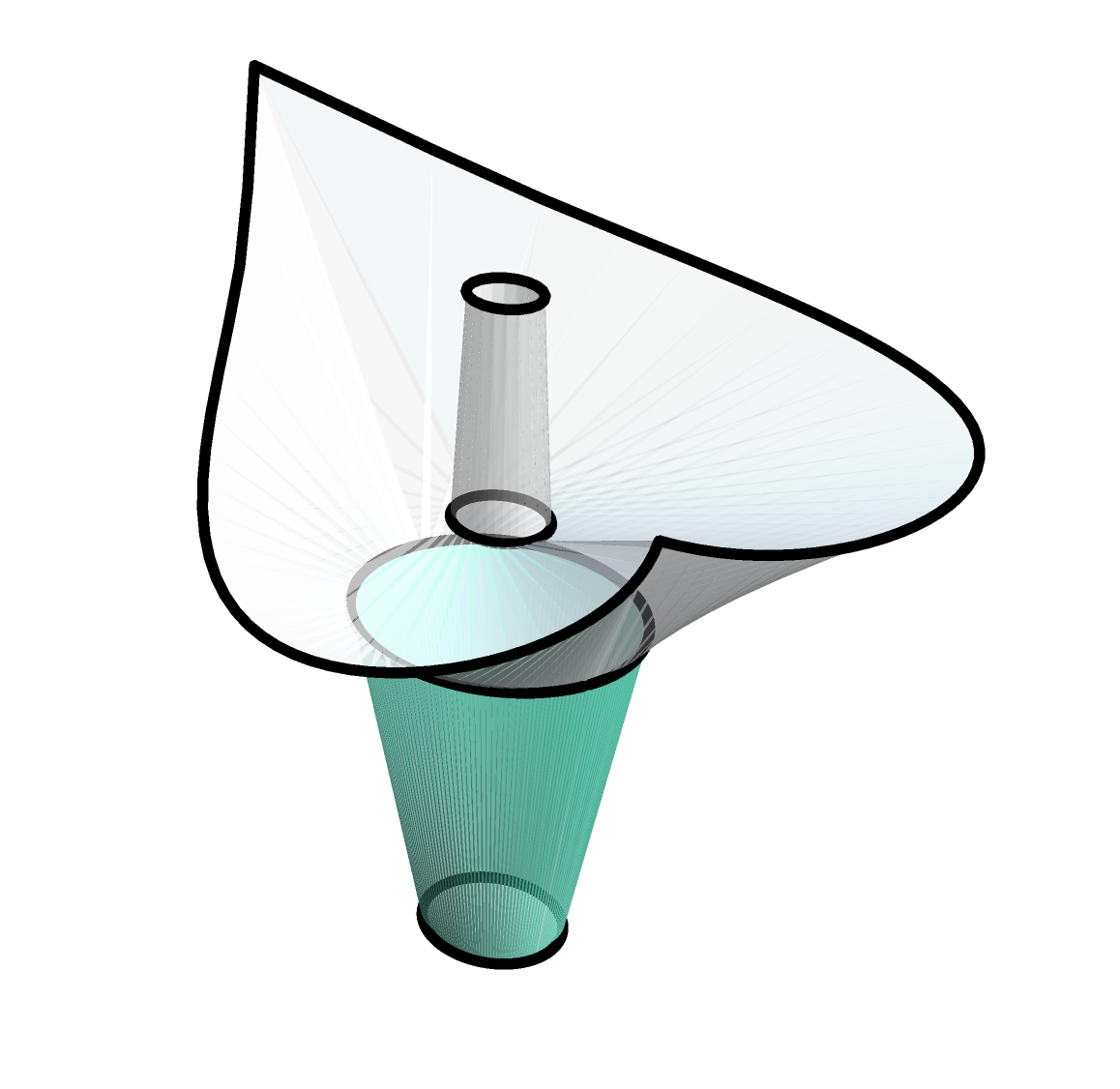}
    \caption{A calla lily-inspired graph with a total of 627 points, including 303 unpinned points. Black thick lines represent layers consisting of triangles.}
    \label{fig:calla-lilly}
\end{figure}

We suggest applying these methods to no more than 1000 points if using only marching methods without stabilization (e.g., recalculating using different directions). We demonstrate how error increases with one example. The shape of the example is inspired by a calla lily, consisting of 303 unpinned points with 327 pinned points, as shown in Fig. \ref{fig:calla-lilly}.

We reconstructed this surface using the linear matrix method under the R platform. The error was calculated as the maximum value in each layer for the Euclidean distance between the target vertex and the reconstructed vertex position vectors. We observed that the error increases by approximately 1000 times when we increase the number of layers.

\begin{table}[ht]
    \centering 
    \begin{tabular}{c c c c c c} 
        \hline\hline
        & Layer\#1 & Layer\#2 & Layer\#3 & Layer\#4 & Layer\#5 \\ [0.5ex] 
        \hline
        Error & N/A & 2.40e-10 & 3.04e-07 & 4.07e-05 & 1.09e-1 \\ 
        [1ex] 
        \hline 
    \end{tabular}
    \caption{Accumulated error for each layer.}
\end{table}

Note that the error inflation is related to how the error is stabilized and the precision level of the program used for the computation.
\subsubsection{Alternative Methods for Shape Reconstruction}
While sequential computation methods (marching methods) can suffer from error accumulation due to sequential matrix calculations, there are several ways to mitigate or avoid these issues. For grid structures that ensure unique realization, we can: Use different marching directions to reduce error accumulation, and employ alternative methods that avoid marching entirely. One such alternative approach uses the complete distance matrix to compute unknown points. For structures with unique realizations, like the strips in Fig.11 (c), we can use Cayley-Menger determinant properties to complete the distance matrix.

\begin{definition}[Cayley-Menger Determinant]\label{CMDef}
For points $P_0, P_1, ..., P_n$ in Euclidean space with pairwise distances $d_{ij} = ||P_i - P_j||$, the Cayley-Menger determinant is defined as:
\begin{equation}
    CM(P_0,...,P_n) = \begin{vmatrix}
    0 & 1 & 1 & \cdots & 1 \\
    1 & 0 & d_{01}^2 & \cdots & d_{0n}^2 \\
    1 & d_{10}^2 & 0 & \cdots & d_{1n}^2 \\
    \vdots & \vdots & \vdots & \ddots & \vdots \\
    1 & d_{n0}^2 & d_{n1}^2 & \cdots & 0
    \end{vmatrix}
\end{equation}
\end{definition}

\begin{theorem}
For points in $\mathbb{R}^d$:
\begin{enumerate}
    \item For $n+1$ points, realizability requires:
        \begin{equation}
(-1)^{k+1}CM(P_0,...,P_k) \geq 0\text{ for }k = 1,...,d            
        \end{equation}
    \item For $n+2$ points, additional condition:
        \begin{equation}
            CM(P_0,...,P_n,P_{n+1}) = 0
        \end{equation}
    \item For $n+3$ points, conditions:
        \begin{align}
            CM(P_0,...,P_n,P_{n+1}) &= 0 \\
            CM(P_0,...,P_n,P_{n+2}) &= 0 \\
            CM(P_0,...,P_n,P_{n+1},P_{n+2}) &= 0
        \end{align}
\end{enumerate}
\end{theorem}
\begin{example}[Computing Missing Distance Using Embedding Conditions]
Consider our structure where: We have 5 points in $\mathbb{R}^3$ with known positions (base vertices); one additional point (blue) connected to 4 base points with known distances; and we seek the distance $d$ between two specific base points (red edge).
 \begin{figure}
\begin{center}    
    \includegraphics[width=0.4 \textwidth]{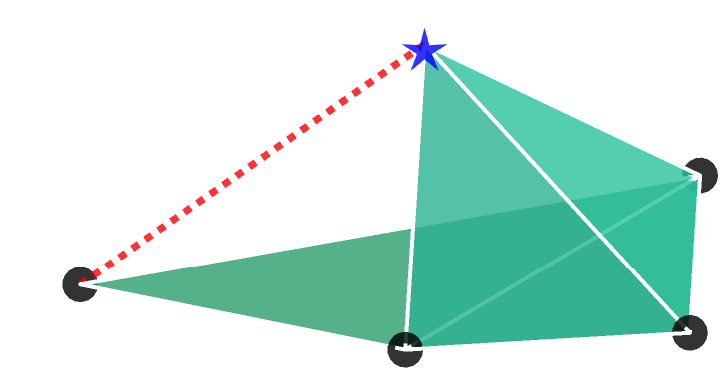}
\caption{{\it Graphical illustration of a missing (red dashed) edge in the grid}: Four pinned points (black dots) with one unknown point (blue star) form a pyramid structure. The white lines on the green faces represent known constraints, while the red dashed edge is unknown. However, this missing edge length can be computed using the Cayley-Menger (CM) matrix.}
\label{fig:missing-edge}
\end{center}
\end{figure}

For each tetrahedron formed with the unknown point, we can write Cayley-Menger conditions:
\begin{align}
    CM(P_0,...,P_4) &= 0 \text{ (first tetrahedron)} \\
    CM(P_0',...,P_4') &= 0 \text{ (second tetrahedron)}
\end{align}

Since these tetrahedra share points and the structure must be embeddable in $\mathbb{R}^3$, we can use the embedding conditions for $n+2$ and $n+3$ points to write additional constraints:
\begin{equation}
    CM(P_0,...,P_5) = 0
\end{equation}

These conditions together determine the possible values for the unknown distance $d$.
\end{example}

This example illustrates how we can fill out the missing edge lengths.
\begin{example}[Pyramid Edge Computation]
For a tetrahedron formed with vertices $P_1$, $P_2$, $P_3$, and $P_5$ (blue point), with known distances: $d_{12} = 1$ (base edge), $d_{23} = 1$ (base edge), $d_{13} = \sqrt{2}$ (diagonal), $d_{15} = \sqrt{3}/2$ (side edge), $d_{25} = \sqrt{3}/2$ (side edge), and $d_{35} = \sqrt{3}/2$ (side edge).
The Cayley-Menger determinant for this tetrahedron is:
\begin{equation}
\begin{vmatrix}
0 & 1 & 1 & 1 & 1 \\
1 & 0 & 1 & 2 & 3/4 \\
1 & 1 & 0 & 1 & 3/4 \\
1 & 2 & 1 & 0 & 3/4 \\
1 & 3/4 & 3/4 & 3/4 & 0
\end{vmatrix} = 0
\end{equation}

The second tetrahedron (with the unknown edge $d$) must also satisfy its Cayley-Menger determinant:
\begin{equation}
\begin{vmatrix}
0 & 1 & 1 & 1 & 1 \\
1 & 0 & d^2 & 1 & 3/4 \\
1 & d^2 & 0 & 1 & 3/4 \\
1 & 1 & 1 & 0 & 3/4 \\
1 & 3/4 & 3/4 & 3/4 & 0
\end{vmatrix} = 0
\end{equation}

By using cofactor expansion along the first row for the second tetrahedron, this expands to a polynomial in $d^2$. The expansion (which can be computed using symbolic mathematics software) leads to:
\begin{equation}
16d^4 - 24d^2 + 9 = 0
\end{equation}

Solving this quadratic in $d^2$:
\begin{align}
16d^4 - 24d^2 + 9 &= 0 \\
d^2 &= \frac{24 \pm \sqrt{576-576}}{32} \\
d^2 &= \frac{24}{32} = \frac{3}{4}
\end{align}

Therefore, $d = \sqrt{3}/2$, confirming our expected result.
\end{example}
Once the distance matrix is completed, coordinates can be reconstructed using the gram matrix method:
\begin{equation}
    G_{ij} = \frac{1}{2}(d_{1i}^2 + d_{1j}^2 - d_{ij}^2)
\end{equation}
where the eigendecomposition of $G$ yields the vertex coordinates.
\bibliography{apssamp}

\end{document}